\documentstyle[12pt]{article}
\newcommand{\gsim}{ \raisebox{-.5ex}{\mbox{$\,\stackrel{>}{\sim}\,$}} }
\newcommand{\lsim}{ \raisebox{-.5ex}{\mbox{$\,\stackrel{<}{\sim}$\,}} }
\normalbaselines
\begin{document}
\baselineskip=15pt
\begin{center}
{\large\bf LINEAR AND NONLINEAR DYNAMICAL CHAOS\footnote{Lectures on the
Intern. Summer School "Nonlinear Dynamics and Chaos", Ljubljana,
Slovenia, 1994}}\\[2mm]
                   Boris Chirikov\\
{\it Budker Institute of Nuclear Physics \\
        630090 Novosibirsk, Russia \\
          chirikov @ inp.nsk.su}\\[5mm]
\end{center} 
\vspace{2mm}
\begin{abstract}
Interrelations between dynamical and statistical laws in physics, on the
one hand, and between the classical and quantum mechanics, on the other hand,
are discussed with emphasis on the new
phenomenon of dynamical chaos.\\
The principal results of the studies into chaos in classical mechanics are
presented in some detail, including the strong local instability and robustness
of the motion, continuity of both the phase space as well as the motion
spectrum, and time reversibility but nonrecurrency of statistical evolution,
within the general picture of chaos as a specific case of dynamical behavior.\\
Analysis of the apparently very deep and challenging contradictions of this 
picture with the quantum principles is given. The quantum view of dynamical
chaos, as an attempt to resolve these contradictions guided by the 
correspondence principle and based upon the characteristic time scales of
quantum evolution, is explained. The picture of the quantum chaos as a new
generic dynamical phenomenon is outlined together with a few other examples
of such a chaos including linear (classical) waves and digital computer.\\
I conclude with discussion of the two fundamental physical problems:
the quantum measurement ($\psi$--collapse), and the causality
principle which both appear to be related to the phenomenon of dynamical
chaos.
\end{abstract}

\newpage

\section{General introduction: statistical properties of dynamical systems}

The main purpose of my lectures is an overview of recent
studies into a new phenomenon (or rather
a whole new field of phenomena) known as the {\it dynamical chaos} both in
classical and especially in quantum mechanics. The concept of dynamical
chaos resolves (or, at least, helps to do so) the two fundamental problems 
in physics and, hence, in all the natural sciences:
\begin{itemize}
\item{} are the dynamical and statistical laws of a different
          nature or one of them, and which one, follows from the other;
\item{} are the classical and quantum mechanics of a different nature or
          the latter is the most universal and general theory currently
          available to describe the whole empirical evidence including
          the classical mechanics as the limiting case.
\end{itemize}

The important part of my philosophy in discussing the problem of chaos 
is the separation of the human from
the natural following Einstein's approach to the science: {\it building up
a model of the real world}. Clearly, the human is also a part of the world,
and moreover the most important one for us as human beings but not as
physicists. The whole phenomenon of life is extremely specific, and one
should not transfer its peculiarities into other fields of natural sciences.

A rather popular nowadays human--oriented philosophy in physics
is the information--based representation of natural laws, particularly,
by substituting the information for entropy (with opposite sign). In the most 
general way such a philosophy was recently presented by Kadomtsev [1].
Such an approach is possible, and it might be done in a selfconsistent way but
one should be very careful to avoid many confusions. In my opinion, the
information is an adequate conception only for the special systems which
actually use and process the information like various automata both natural
(living systems) as well as man--made ones. In this case the information
becomes a physical notion rather than a human view of natural phenomena. 
The same is also true in the theory of measurement which is again a very
specific physical process, the basic one in our studies of the Nature but
still not a typical one for the Nature itself. This is crucially important in
quantum mechanics as will be discussed in some detail below (Lectures 4 and 5).

One of the major implications from the studies in dynamical chaos is the
conception of statistical laws as an intrinsic part of dynamics without any
additional statistical hypotheses (for the current state of the theory see,
e.g., Ref.[2] and recent Collection of papers [3] as well as the Introduction
to this Collection [4]). This basic idea can be traced back to Poincare [5]
and Hadamard [6], and even to Maxwell [7], the principal condition for 
dynamical chaos being strong local instability of motion. In this
picture the statistical laws are considered as {\it secondary} with respect to
more fundamental and general {\it primary} dynamical laws.

Surprisingly, the opposite is also true!

Namely, under certain conditions the dynamical laws were found to be completely
contained in the statistical ones. Nowadays this is called 'synergetics' [8]
but the principal idea goes back to Jeans [9] who discovered the instability
of gravitating gas (a typical example of statistical system) which is the
basic mechanism for the formation of galaxies and stars in the modern 
cosmology, and eventually the Solar system, a classical example of dynamical
system. In this case the resulting dynamical laws proved to be secondary
with respect to the primary statistical laws which include the former. 

Thus, the whole picture can be represented as a chain of dynamical--statistical
inclusions:
$$ ...?...\,\fbox{$ D\,\supset\,S $}
   \,\supset\,D\,\supset\,S\,...?... \eqno (1.1)
$$
Both ends of this chain, if any, remain unclear. So far the most fundamental
(elementary) laws of physics seem to be dynamical (see, however, discussion
of quantum measurement in Lectures 4 and 5). This is why I begin the chain 
(1.1) with some primary dynamical laws.

The strict inclusion on each step of the chain has a very important consequence
allowing for the so--called numerical experiments, or computer simulation,
of a broad range of natural processes. As a matter of fact the former (not
laboratory experiments) are now the main source of new information in the
studies of the secondary laws for both dynamical chaos and synergetics.
This might be called {\it the third way of cognition}, in addition to 
laboratory experiments and theoretical analysis.

In what follows I restrict myself to the discussion of just a single ring
of the chain as marked in Eq.(1.1). Here I will consider the dynamical chaos
separately in classical and quantum mechanics. In the former case the chaos
explains the origin and mechanism of random processes in the Nature (within
the classical approximation). Moreover, that deterministic randomness may
occur (and is typical as a matter of fact) even in minimal number of freedoms
$N>1$ (for Hamiltonian systems), thus enormously expanding the domain for
application of the powerful methods of statistical analysis. 
The latter provides a rather simple (see, however, Lecture 3) description
of the essential features for the otherwise highly intricate dynamical
motion.

In quantum mechanics the whole situation is much more tricky, and still
remains rather controversial. Here we encounter an intricate tangle of
various apparent contradictions between the correspondence principle,
classical chaotic behavior, and the very foundations of quantum physics.
This will be the main topic of my discussions in Lecture 4.

One way to untangle this tangle is the new general conception -
{\it pseudochaos}, of which the quantum chaos is the most important example. 
Another interesting example is the digital computer, also very important 
in view of broad application of numerical experiments in the studies of 
dynamical systems. On the other hand, the pseudochaos in computer will
hopefully help to understand quantum pseudochaos and to accept it as a sort
of {\it chaos} rather than of a regular motion as many researchers, even in
this field, still do believe. 

The new and surprising phenomenon of dynamical chaos, especially in quantum
mechanics, holds out new hopes for eventually solving some old, long--standing,
fundamental problems in physics. In Lecture 5 I will briefly discuss two of
them:
\begin{itemize}
\item{} causality principle (time ordering of cause and effect), and
\item{} $\psi$--collapse in the quantum measurement.
\end{itemize}

The conception of dynamical chaos I am going to present here, which is not
common as yet, was the result of the long--term Siberian--Italian (SI)
collaboration including Giulio Casati and Italo Guarneri
(Como), and Felix Izrailev and Dima Shepelyansky (Novosibirsk) with whom I
share the responsibility for our joint scientific results and their conceptual
interpretation.

\section{Chaos in classical mechanics: dynamical complexity}

The classical dynamical chaos, as a part of classical mechanics, was 
historically the first to have been studied simply because in the time of
Boltzmann, Maxwell, Poincare and other founders of statistical mechanics
the quantum mechanics did not exist. No doubt, the general mathematical theory
of dynamical systems, including the ergodic theory as its modern part
describing various statistical properties of the motion, has arisen from
(and is still conceptually based on) the classical mechanics [10]. Yet, upon
construction, it is not necessarily restricted to the latter and can be applied
to a much broader class of dynamical phenomena, for example, in the quantum
mechanics (Lecture 4).

\subsection{Dynamical systems}

In classical mechanics dynamical system means an object whose motion in some
{\it dynamical space} is {\it completely} determined by a given interaction
and the {\it initial conditions}. Hence, a synonym {\it deterministic system}.
The motion of such a system can be described in two seemingly different ways
which, however, prove to be essentially equivalent. 

The first one are the {\it motion equations} of the form
$$ \frac{d{\bf x}}{dt}\,=\,{\bf v}({\bf x},\,t) \eqno (2.1)
$$
which always have a unique solution
$$ {\bf x}\,=\,{\bf x}(t,\,{\bf x_0}) \eqno (2.2) 
$$
Here ${\bf x}$ is a finite--dimensional vector in the dynamical space and 
${\bf x_0}$ the
initial conditions (${\bf x_0}={\bf x}(0)$). A possible explicit time--dependence in
r.h.s. of Eq.(2.1) is assumed to be regular, e.g., periodic one or, at least,
that with discrete spectrum.

The most important feature of dynamical systems is the 
{\it absence of any random parameters or any noise} in the motion equations.
Particularly, for this reason I will consider a special class of dynamical
systems, the so--called {\it Hamiltonian (nondissipative) systems}, which are
most fundamental in physics. 

{\it Dissipative systems}, being very important in many applications, are
neither fundamental (because the dissipation is introduced via a crude
approximation of the very complicated interaction with some 'heat bath')
nor purely dynamical in view of principally inevitable random noise in the
heat bath (fluctuation--dissipation theorem). In a more accurate and 
natural way the dissipative systems can be described in the frames of the
secondary dynamics ($S\supset D$ inclusion in Eq.(1.1)) when both dissipation
and fluctuations are present from the beginning in the primary statistical
laws. 

A purely dynamical system is necessarily the {\it closed} one which is the
main object in the fundamental physics. Thus, any coupling to the environment
is completely neglected. I will come back to this important question below.

In Hamiltonian mechanics the dynamical space, called {\it phase space}, is
even--dimensional one composed of $N$ pairs of canonically conjugated 
'coordinates'
and 'momenta', each pair corresponding to one freedom of motion.

In the problem of dynamical chaos the initial conditions play a special role:
they completely determine a particular trajectory, for a given interaction,
or a particular realization of dynamical process which may happen to be a
very specific, nontypical, one. To get rid of such singularities another
description is useful, namely, the Liouville partial differential equation
for the {\it phase space density}, or distribution function $f({\bf x},\,t)$:
$$ \frac{\partial f}{\partial t}\,=\,\hat{L}\,f \eqno (2.3)
$$
with the solution
$$ f\,=\,f({\bf x},\,t;\,f_0({\bf x})) \eqno (2.4)
$$
Here $\hat{L}$ is {\it linear} differential operator, and $f_0({\bf x})=
f({\bf x},\,0)$
the initial density. For any smooth $f_0$ this description provides the generic
behavior of dynamical system via a continuum of trajectories. In special case
$f_0=\delta ({\bf x}-{\bf x_0})$ the density describes a single trajectory 
like the motion equations (2.1). Notice that even in this limiting case
Eq.(2.3) is linear with respect to the dynamical variable $f$.

In any case the phase space itself is assumed to be {\it continuous} which is
the most important feature of the classical picture of motion and the main
obstacle in the understanding of quantum chaos.

\subsection{Dynamical chaos}

Dynamical chaos can be characterized in terms of both the individual 
trajectories and the trajectory ensembles, or phase density. Almost all
trajectories of a chaotic system are in a sense most complicated 
({\it unpredictable} from observation of any preceding motion).
Exceptional, e.g., periodic trajectories form a set of zero
invariant measure, yet it might be everywhere dense.

An appropriate notion in the theory of chaos is {\it symbolic trajectory}
first introduced by Hadamard [6]. The theory of symbolic dynamics was
developed further in Refs.[11 -- 13]. Symbolic trajectory is a projection
of the true (exact) trajectory on a discrete partition of the phase space
at discrete instants of time $t_n$, e.g., such that $t_{n+1}-t_n=T$ fixed.
In other words, to obtain a symbolic trajectory we first turn from the motion
differential equations (2.1) to the difference equations over a certain
time interval $T$:
$$ {\bf x}(t_{n+1})\,\equiv\,{\bf x}_{n+1}\,=\,M({\bf x}_n,\,t_n) \eqno (2.5)
$$
These are usually called {\it mapping} or {\it map}: ${\bf x}_n\to 
{\bf x}_{n+1}$.
Then, while running (theoretically) {\it exact} trajectory we record
each ${\bf x}_n$ to a {\it finite} accuracy: ${\bf x}_n\approx m_n$. 
For a finite partition
each $m_n$ can be chosen integer. Hence, the whole infinite symbolic
trajectory
$$ \sigma\,\equiv\,...m_{-n}...m_{-1}m_0m_1...m_n...\,=S({\bf x_0};\,T) 
   \eqno (2.6)
$$
can be represented by a {\it single} number $\sigma$ which is generally
irrational, and which is some function of the {\it exact} initial conditions.
The symbolic trajectory may be also called {\it coarse--grained trajectory}.
I remind that the latter is a {\it projection} of (not substitution for)
the exact trajectory to represent in compact form the global dynamical
behavior without unimportant microdetails.

A remarkable property of chaotic dynamics is in that the set of its symbolic
trajectories is {\it complete} that is it actually contains all possible
sequences (2.6). Apparently, this is related to continuity of function
$S({\bf x_0})$ (2.6). On the opposite, for a regular motion this function is 
everywhere discontinuous.

In a similar way the {\it coarse--grained} phase density $\overline{f}
(m_n,\,t)$ is introduced, in addition to exact, or {\it fine--grained}
density, which is also a projection of the latter on some partition of the
phase space. 

The coarse--grained density represents the global dynamical behavior,
particularly, the most important process of {\it statistical relaxation},
for chaotic motion, to some {\it steady state} $f_s(m_n)$ (statistical
equilibrium) independent of initial $f_0({\bf x})$ if the steady state is 
{\it stable}. Otherwise, synergetics comes into play giving rise to a
secondary dynamics (Lecture 1). As the relaxation is aperiodic process 
the spectrum of
chaotic motion is {\it continuous} which is another obstacle for the theory
of quantum chaos.

The relaxation is one of the characteristic properties of statistical
behavior. Another one are {\it fluctuations}. Chaotic motion is a generator
of noise which is purely {\it intrinsic} by definition of the dynamical
system. Such a noise is a particular manifestation of the complicated dynamics
as represented by the symbolic trajectories or by the difference
$$ f({\bf x},\,t)\,-\,\overline{f}(m_n,\,t)\,\equiv\,\tilde{f}({\bf x},\,t) 
   \eqno (2.7)
$$

The relaxation $\overline{f}\to f_s$, apparently asymmetric with respect to
time reversal $t\to -t$, had given rise to a long--standing misconception
of the notorious {\it time arrow}. Even now some very complicated mathematical
constructions are still being erected (see, e.g., Refs.[14]) in attempts to
extract somehow statistical irreversibility from the reversible mechanics.
This is especially surprising as such 'irreversibility' is based on the
separation of the phase density into two parts similar to Eq.(2.7).
In fact, the time direction is fixed by the additional statistical condition
imposed on initial $f_0$ which is equivalent also to the 'causality condition'
(see Lecture 5).

In the theory of dynamical chaos there is no such problem. The answer turns
out to be conceptual rather than physical: one should separate two similar
but different notions, {\it reversibility} and {\it recurrency}. The exact
density $f({\bf x},\,t)$ is always {\it time--reversible} but 
{\it nonrecurrent}
for chaotic motion that is it will never come back to the initial 
$f_0({\bf x})$
in {\it both} directions of time $t\to\pm\infty$. In other words, the 
relaxation, also present in $f$, is time--symmetric. The projection of $f$,
coarse--grained $\overline{f}$, which is both nonrecurrent and irreversible,
emphasizes nonrecurrency of the exact solution. The apparent violation of 
the statistical relaxation upon time reversal, as described by the exact 
$f({\bf x},\,t)$,
represents in fact the growth of a big fluctuation which eventually will be
followed by the same relaxation in the opposite direction of time. This
apparently surprising symmetry of the statistical behavior was discovered 
long ago by Kolmogorov [15]. Another manifestation of that symmetry is
the well--known principle of detailed balancing (for discussion see, e.g.,
Ref.[24]).

One can say that instead of imaginary time
arrow there exists the {\it process arrow} pointing always to the steady state.
The following simple example would help, perhaps, to overcome this 
conceptual difficulty. Consider the hyperbolic one--dimensional (1D) motion:
$$ x(t)\,=\,a\cdot\exp{(\Lambda t)}\,+\,b\cdot\exp{(-\Lambda t)} \eqno (2.8)
$$
which is obviously time--reversible, yet remains {\it unstable} in both
directions of time ($t\to\pm\infty$). Besides its immediate appealing this 
example is closely related to the mechanism of chaos which is the motion
instability. Another example of time--reversible chaos will be given in
Lecture 3.

\subsection{Instability and chaos: dynamical complexity}

Local instability of motion responsible for a very complicated dynamical
behavior is described by the {\it linearized equations}:
$$ \frac{d{\bf u}}{dt}\,=\,{\bf u}\cdot\frac{\partial{\bf v}({\bf x^0}(t),
   \,t)}{\partial{\bf x}} \eqno (2.9)
$$
Here ${\bf x^0}(t)$ is a reference trajectory satisfying Eq.(2.1), and
${\bf u}={\bf x}(t)-{\bf x^0}(t)$ the deviation of a close trajectory 
${\bf x}(t)$. At average, the solution of Eq.(2.9) has a form
$$ |{\bf u}|\,\sim\,\exp{(\Lambda t)} \eqno (2.10)
$$
where $\Lambda$ is called {\it Lyapunov's exponent}. The motion is 
(exponentially) unstable if $\Lambda >0$. In the Hamiltonian system of $N$
freedoms there are $2N$ Lyapunov's exponents satisfying the condition
$\sum\,\Lambda =0$. The partial sum of all positive exponents $\Lambda_+ >0$
$$ h\,=\,\sum\,\Lambda_+ \eqno (2.11)
$$
is called (dynamical) {\it metric entropy}. Notice that it has the dimensions
of frequency and characterises the instability rate.

The motion instability is only a necessary but not sufficient condition for
chaos. Another important condition is {\it boundedness} of the motion, or
its oscillatory (in a broad sense) character. The chaos is produced by the
combination of these two conditions (also called stretching and folding).
Let us again consider an elementary example of 1D map
$$ x_{n+1}\,=\,2\,x_n \quad mod\ 1 \eqno (2.12)
$$
where operation $mod\ 1$ restricts (folds) $x$ to the interval (0,1). This is 
not a Hamiltonian system but it can be interpreted as a 'half' of that,
namely, as the dynamics of the oscillation phase. This motion is unstable
with $\Lambda =\ln{2}$ because the linearized equation is the same except
taking the fractional part ($mod\ 1$). The explicit solution for both reads
$$ 
\begin{array}{ll}
u_n\,=\,2^n\,u_0 & \\
 x_n\,=\,2^n\,x_0 & mod\ 1 
\end{array} 
\eqno (2.13)
$$
The first (linearized) motion is unbounded like Hamiltonian hyperbolic 
motion (2.8) and perfectly regular. The second one is not only unstable but
also chaotic just because of the additional operation $mod\ 1$ which makes
the motion bounded, and which mixes up the points within a finite interval.

The combination of two above conditions for chaos -- exponential instability
and boundedness -- requires the {\it motion equations} to be {\it nonlinear}.
In the latter example (2.12) nonlinearity is provided by the operation
$mod\ 1$. However, Liouville's Eq.(2.3) for the phase density $f$ is always
{\it linear}. Hence, the local stability of $f$ that is the variation for a
small deviation $\delta f=f-f^0$ is described by the same Liouville's
Eq.(2.3). The motion exponential instability ($\Lambda =\pm\Lambda_{\pm}
>0$) results then in the contraction of the domain
occupied by the initial phase density. If the simultaneous stretching in
another direction is bounded (owing to nonlinearity of the motion, not
Liouville's, equation) the exponentially long domain of conserving volume
fills up the whole phase space region allowed by the exact motion integrals,
e.g., the whole energy surface of a conservative system. Eventually,
coarse--grained density $\overline{f}$ approaches 
a homogeneous steady state $f_s$
while the exact density $f$ keeps fluctuating with a characteristic wave
length exponentially decreasing in time. In other words, we may say that
in Liouville's description the phase space density evolution is exponentially
unstable in the wave number (vector) ${\bf k}$ of $f({\bf x})$ rather than in
$f({\bf x})$ itself. Notice that for a Hamiltonian system vector ${\bf x}$
includes momenta as well.

We may look at the above example (2.12) from a different viewpoint. 
Let us express
initial $x_0$ in the binary code as the sequence of two symbols, 0 and 1,
and let us make the partition of unit $x$ interval also in two equal halves
marked by the same symbols. Then, the symbolic trajectory will simply repeat
$x_0$ that is Eq.(2.6) takes the form
$$ \sigma\,=\,x_0 \eqno (2.14)
$$
It implies that, as time goes on, the global motion will eventually depend on
ever diminishing details of the initial conditions. In other words, when we 
formally fix {\it exact} $x_0$ we 'supply' the system with infinite complexity
which is coming up due to the strong motion instability. Still another 
interpretation is in that the exact $x_0$ is the source of {\it intrinsic 
noise} amplified by the instability. For this noise to be {\it stationary}
the string of $x_0$ digits has to be infinite which is only possible in
{\it continuous} phase space. 

A nontrivial part of this picture of chaos is in that the instability must be
{\it exponential} while a power--law instability is insufficient for chaos.
For example, linear instability ($|{\bf u}|\sim t$) is a generic property of 
perfectly regular motion of the completely integrable system whose motion
equations are {\it nonlinear} and, hence, whose oscillation frequencies
depend on initial conditions [16, 17]. The character of motion for a faster
power--law
instability ($|{\bf u}|\sim t^{\alpha},\ \alpha >1$) is unknown.

On the other hand, the exponential instability ($h>0$) is not invariant with
respect to the change of time variable [4] (in this respect the only
invariant statistical property is ergodicity [10]). A possible resolution of
this difficulty is in that the proper characteristic of motion instability,
important for dynamical chaos, should be taken with respect to the oscillation
phases whose dynamics determines the nature of motion. It implies that the 
proper time variable must go proportionally to the phases so that the 
oscillations become stationary [4].
A simple example is harmonic oscillation with frequency $\omega$
recorded at the instances of time $t_n=2^nt_0$. Then, oscillation phase
$x=\omega t/2\pi$ obeys map (2.12) which is chaotic. Clearly, the
origin of chaos here is not in the dynamical system but in the recording
procedure (random $t_0$). Now, if $\omega$ is a parameter (linear
oscillator), then the oscillation is exponentially unstable (in new time $n$) 
but only
with respect to the change of parameter $\omega$, not of the initial
$x_0$ ($x\to x+x_0$). In a slightly 'camouflaged' way essentially the
same effect was considered in Ref.[56] with far--reaching conclusions
for the quantum chaos (Lecture 4).

Rigorous results concerning the relation between instability and chaos 
are concentrated in the Alekseev - Brudno theorem [13] (see also Refs.[4, 18])
which states that the complexity per unit time of almost any symbolic 
trajectory is asymptotically equal to the metric entropy:
$$ \frac{C(t)}{|t|}\,\to\,h\,, \qquad |t|\,\to\,\infty \eqno (2.15)
$$
Here $C(t)$ is the so--called algorithmic complexity, or in more familiar 
terms, the information associated with a trajectory segment of length $|t|$.

The transition time from the dynamical to statistical behavior according to 
Eq.(2.15) depends on the partition of the phase space, namely, on the size of
a cell $\mu$ which is inversely proportional to the biggest integer
$M\geq m_n$ in symbolic trajectory (2.6). The transition is controlled by the
{\it randomness parameter} [19]:
$$ r\,=\,\frac{h\,|t|}{\ln{M}}\,\sim\,\frac{|t|}{t_r} \eqno (2.16)
$$
where $t_r$ is the {\it dynamical time scale}. As both $|t|,\,M\to\infty$ we
have a somewhat confusing situation, typical in the theory of dynamical chaos,
when two limits do not commute:
$$ M\,\to\,\infty ,\,|t|\,\to\,\infty\ \neq\ |t|\,\to\,\infty ,\,M\,\to\,\infty
   \eqno (2.17)
$$
For the left order ($M\to\infty$ first) parameter $r\to 0$, and we have
{\it temporary determinism} ($|t|\lsim t_r$), while for the right order
$r\to\infty$, and we arrive at the {\it asymptotic randomness} 
($|t|\gsim t_r$). 

Instead of the above double limit we may consider the {\it conditional limit}
$$ |t|,\ M\,\to\,\infty , \qquad r\,=\,const \eqno (2.18)
$$
which is also a useful method in the theory of chaotic processes. Particularly, 
for $r\lsim 1$ strong dynamical correlations persist in a symbolic trajectory
which allows for the prediction of trajectory from a finite--accuracy 
observation. This is no longer the case for $r\gsim 1$ when only statistical
description is possible. Nevertheless, the motion equations can still be used to
completely derive all the statistical properties without any {\it ad hoc}
hypotheses. Here the exact trajectory {\it does exist} as well but becomes
the Kantian {\it thing--in--itself} which can be only observed but neither
predicted nor reproduced in any other way. 

The mathematical origin of this peculiar property goes back to the famous
G\"odel theorem [20] which states (in modern formulation) that {\it most}
theorems in a given mathematical system are unprovable, and which forms the
basis of contemporary mathematical logic as well as of the algorithmic
theory of dynamical systems (see Ref.[21] for detailed explanation
and interesting applications of this relatively less known mathematical
achievement). A particular corollary, directly related to symbolic trajectories
(2.6), is that almost all real numbers are uncomputable by any finite
algorithm. Besides rational numbers some irrationals like $\pi$ or $e$ are
also known to be computable. Hence, their total complexity, e.g., $C(\pi )$
is finite, and the complexity per digit is zero (cf. Eq.(2.15)).

The main object of my discussion here, as well as of the whole physics, is a
closed system which requires neglecting the external perturbations. However,
in case of strong motion instability this is no longer possible, at least,
dynamically. What is the impact of a weak perturbation on the statistical
properties of a chaotic system? The rigorous answer was given by the
robustness theorem due to Anosov [22]: not only statistical properties remain
unchanged but, moreover, the trajectories get only slightly deformed providing
(and due to) the same strong motion instability. The explanation of this 
striking peculiarity is in that the trajectories are simply transposed and,
moreover, the less the stronger is instability. 

In conclusion let me make a very general remark, far beyond the particular
problem of chaotic dynamics (see also Ref.[89]). 
According to the Alekseev - Brudno theorem (2.15)
the source of stationary (new) information is always chaotic. Assuming farther
that any creative activity, science including, supposed to be such a source
we come to an interesting conclusion that any such activity has to be (partly!)
chaotic. This is the creative side of the chaos.

\section{Chaos in classical mechanics: statistical complexity}

The theory of dynamical chaos does not need any statistical hypotheses, nor
does it allow for arbitrary ones. Everything is to be deduced from the 
dynamical equations. Sometimes the statistical properties turn out to be quite
simple and familiar [2,23]. This is usually the case if the chaotic motion is 
also ergodic (on the energy surface).
However, quite often, and even typically for a 
few--freedom chaos, the phase space is divided, and the chaotic component of 
the motion has a very complicated structure which results in a high complexity
not only of individual trajectories (Lecture 2) but also of the statistical
picture of the motion. Before to proceed further
let us consider a few simple examples.

\subsection{Simple physical examples of dynamical chaos}

In these Lectures I restrict myself to finite--dimensional systems where the 
peculiarities of dynamical chaos are most clear (see Lecture 4 for some
brief remarks on infinite systems). Consider now a few examples of chaos
in minimal dimensionality. In a conservative system of one freedom ($N=1$)
chaos is impossible. Such a system is {\it completely integrable} since
there is one motion integral, the energy, per one freedom. The motion is
periodic that is perfectly regular. The solution (2.2) of motion equations
(2.1) is explicitly expressed in the standard way as the integral of the
Hamiltonian. Chaos requires at least two freedoms (for conservative systems).
For a regular (quasiperiodic) motion two independent (commuting) and isolating
(single--valued)
integrals would be necessary which is not always the case. At this point
I would like to mention a rather widespread confusion that any motion equations
possess $2N$ integrals, the initial conditions (see Eq.(2.2)). This is 
certainly true but those integrals are nonisolating, in fact they might be
infinitely many--valued. In the latter case, the trajectory is not restricted
to an {\it invariant surface} of lower dimensions, and may be even 
{\it ergodic} that is occupy the whole energy surface.
Let me mention also that there are some minor differences between several
possible definitions of the (complete) integrability. One is based on the
motion integrals in some particular dynamical space. Another one (more narrow)
corresponds to a stronger condition of the existing of the motion integrals
in {\it action--angle} variables\footnote{I denote actions by $n$ having in
mind the subsequent quantization in Lecture 4 when they become integers if
 $\hbar =1$.}
${\bf n},\ {\bf \phi}$. In this case the 
invariant surface of the completely integrable system is an $N$--dimensional 
{\it torus}. Below I will asume the latter definition of integrability.

In case of time--dependent Hamiltonian $H(n,\phi ,t)$ the chaos is 
possible even in one freedom. This is because such a system is equivalent to
the conservative 2--freedom one in the {\it extended phase space}
$n_1,\,n_2,\,\phi_1,\,\phi_2$ with a new Hamiltonian [2]
$$ \overline{H}(n_1,\,n_2,\,\phi_1,\,\phi_2)\,=\,H(n_1,\phi_1,\phi_2/
   \Omega )\,+\,\Omega\,n_2\,=\,0  \eqno (3.1)
$$
where $n_2=-H/\Omega$, and $\phi_2=\Omega t$
(for periodic time--dependence of frequency $\Omega$). In this case one speaks 
sometimes on
one--and--a--half freedoms ($N=1.5$) as the time dependence is fixed.

\subsubsection{Charged particle confinement in adiabatic magnetic traps}

This is Budker's problem [24, 25], very important in the studies of controlled
nuclear fusion. A simple model of two freedoms (axisymmetric magnetic 
field) is described by the Hamiltonian:
$$ H\,=\,\frac{p^2}{2}\,+\,\frac{(1\,+\,x^2)\,y^2}{2} \eqno (3.2)
$$
Here magnetic field $B=\sqrt{1+x^2};\ p^2=\dot{x}^2+\dot{y}^2;\ x$ describes
the motion along magnetic line, and $y$ does so across the line (a projection
of Larmor's rotation). 

In Ref.[24] a slightly different model with 'potential energy' (which is
actually the transverse part of particle's kinetic energy)
$U=(1+x^2)^2\,y^2/2$ was considered in detail. Here I chose model (3.2)
to apply the results below to a completely different physical system.

Assume the adiabaticity parameter
$$
  \lambda\,=\,\frac{1}{v_0}\,\sim\,\omega_y(0)\cdot\tau_x\,\gg 1 \eqno (3.3)
$$
where $v_0$ is the full particle velocity ($H=v_0^2/2$), 
$\omega_y(x)=\sqrt{1+x^2}$ stands for the frequency of transverse oscillation,
and $\tau_x\sim 1/v_0$ is a characteristic time for crossing the magnetic
field minimum at $x=0$. 
Under this condition both actions, $n_x$ and $n_y$, which
are also adiabatic invariants, are approximately conserved. This would
imply bounded $x$--oscillations that is the confinement of a particle in
magnetic trap. However, the adiabatic invariant is only an approximate
motion integral, and Budker's problem is the evaluation of a long--term
variation of that, if any, which would result in a leakage of particles
out of the trap. 

The unperturbed (adiabatic) Hamiltonian for model (3.2) is defined by
$n_y=\omega_ya_y^2/2=const$ with $y=a_y\cos{\phi_y}$, and reads:
$$
  H_0\,=\,\frac{p_x^2}{2}\,+\,n_y\,\omega_y(x)\,\approx\,
  \left(\frac{3\pi}{4\sqrt{2}}\,n_x\,n_y\right)^{2/3}\,\approx\,
  const \eqno (3.4)
$$
Consider the case of large $x$--oscillation, with amplitude
$a_x=H_0/n_y\gg 1$. Then, the frequency
$$
  \omega_x\,\approx\,\frac{\pi}{2}\,\sqrt{\frac{n_y}{2a_x}}\,=\,
  \frac{\pi}{2\sqrt{2}}\,\frac{n_y}{\sqrt{H_0}}\,=\,\frac{\partial H_0}
  {\partial n_x} \eqno (3.5)
$$
Hence, last expression for $H_0$ in Eq.(3.4), and
$$
  <\omega_y>\,=\,\frac{\partial H_0}{\partial n_y}\,=\,\frac{2}{3}\,
  \frac{H_0}{n_y} \eqno (3.6)
$$
where the brackets denote averaging over $x$--oscillation.

Now, the central part of the problem -- evaluation of $n_y$ variation from
the equation:
$$
  \frac{\dot{n_y}}{n_y}\,=\,\frac{d}{dt}\,\ln{n_y}\,=\,
  \frac{x\,\dot{x}}{1\,+\,x^2}\cdot\cos{2\phi} \eqno (3.7)
$$
where we drop sub $y$. This equation can be derived either via canonical
transformation of the original Hamiltonian (3.2) to action--angle variables
or directly from the exact motion equations.

Under adiabatic condition (3.3) $n_y$ variation is exponentially small in
parameter $\lambda$. So, this part of the problem is essentially
{\it nonperturbative} that is it cannot be solved using conventional
perturbation techniques of the expanding in an asymptotic power series
in small parameter $1/\lambda$. Instead, a new perturbation parameter
should be introduced absorbing nonadiabatic exponential. To this end we
integrate Eq.(3.7) over half--period of $x$--oscillation substituting
in r.h.s the unperturbed solution.

The integration is performed in the complex plane of phase $\phi$
$$
  \Delta\,\ln{n_y}\,=\,{\rm Re}\,\int\frac{x\,\dot{x}}{1\,+\,x^2}
  \cdot\exp{(2i\phi )}\,d\phi\,=\,\epsilon_a\cdot\sin{2\phi_0} \eqno (3.8)
$$
around the cut from the branch point at $x=x_p=i$ and $\phi =\phi_p$
to infinity. Here
$$
  \phi_p\,=\,\phi_0\,+\,\int_0^{x_p}\,\frac{\omega_y\,dx}{\dot{x}}\,
  \approx\,\phi_0\,+\,i\,\frac{\pi}{4v_0} \eqno (3.9)
$$
$\dot{x}\approx v_0$, and $\phi_0$ is the phase value at $x=0$. For 
$\lambda\gg 1$ the integral can be reduced to $\Gamma$--function, and we
obtain for the amplitude in Eq.(3.8)
$$
  \epsilon_a\,\approx\,\frac{2\pi}{3}\,\exp{\left(-\,\frac{\pi}{2}\,
  \lambda\right)} \eqno (3.10)
$$
This is the required perturbation parameter.

Now we can derive a map describing the particle motion over many
$x$--oscillations. Beside Eq.(3.8) we need another one for phase $\varphi =
2\phi_0$. From Eqs.(3.5) and (3.6) we have
$$
  \Delta\varphi\,=\,2\pi\,\frac{<\omega_y>}{\omega_x}\,=\,\frac{4}{3}\,
  \frac{v_0^3}{n_y^2}\,=\,G(P) \eqno (3.11)
$$
where a new variable $P=\ln{n_y}$ is introduced. Now, from Eqs.(3.8) and
(3.11) we arrive at the map $(P,\,\varphi )\,\to\,(\overline{P},\,
\overline{\varphi})$ over half--period of $x$--oscillation:
$$
\begin{array}{l}
  \overline{P}\,=\,P\,+\,\epsilon_a\cdot\sin{\varphi}\\
  \overline{\varphi}\,=\,\varphi\,+\,G(\overline{P})    
\end{array} \eqno (3.12)
$$
In the second equation the new value of momentum ($\overline{P}$) is
substituted which determines the change in phase $\varphi$ up to the next
crossing the plane $x=0$ where the first equation operates. As $v_0=const$
is the exact motion integral the map (3.12) is canonical, particularly
preserving the phase plane area $d\Gamma_2=dP\cdot d\varphi$. 

The map describes the global dynamics of the model and
is relatively simple for further analysis both in numerical
experiments as well as by means of asymptotic perturbation series in the new
small parameter $\epsilon_a$. It can be still simplified
by linearizing the second Eq.(3.12) around a resonance at $P=P_r$ such that
$G(P_r)=2\pi r$ with any integer $r$. Upon dropping the latter term the map
is reduced to the so--called {\it standard map} which (in standard notations)
reads[23]:
$$
\begin{array}{l}
  \overline{p}\,=\,p\,+\,k\cdot\sin{\varphi}\\
  \overline{\varphi}\,=\,\varphi\,+\,T\cdot\overline{p}    
\end{array} \eqno (3.13)
$$
where $p=P-P_r,\ k=\epsilon_a$, and new parameter
$$
  T\,=\,\frac{dG(P)}{dP}\,=\,-\,\frac{8}{3}\,v_0^3\,\exp{(-2P_r)} \eqno (3.14)
$$

The term 'standard' emphasizes a universal character of the map to which
many (but, of course, not all) various physical models can be reduced as we
shall see right below. Both maps, (3.12) and (3.13), can be formally
considered as describing one--freedom system driven by the periodic external
perturbation in the form of short $\delta$--pulses. Hence, a nickname
'kicked rotator' for model (3.13). Yet, contrary to a common belief, the
map can describe also a {\it conservative} system as is the case in our
example. Then, it is called the {\it Poincare map} [2].

Unlike global map (3.12) the standard map describes the dynamics 
locally in momentum, e.g., $P$ for Eq.(3.12). This dynamics is determined
by a single parameter
$$
  K\,=\,|k\,T|\,=\,\frac{16\pi}{9\lambda^3}\,\exp{\left(-\,\frac{\pi\lambda}
  {2}\,-\,2P\right)}\,>\,1 \eqno (3.15)
$$
The latter inequality determines the region of chaotic motion in parameter
$K$ for the standard map, and that in phase space for model (3.2) [23].
In the latter case the chaos condition becomes:
$$
  n_y^2\,<\,\frac{16\pi}{9\lambda^3}\,{\rm e}^{-\,\frac{\pi\lambda}
  {2}} \quad {\rm or} \quad \beta_0\,<\,\left(\frac{64\pi}{9}\right)^{1/4}\,
  \lambda^{1/4}\,{\rm e}^{-\,\frac{\pi\lambda}{8}}\,=\,\beta_b \eqno (3.16)
$$
where $\beta_0\approx v_y/v_0\ll 1$ is the so--called pitch--angle at $x=0$.
The second inequality (3.16) determines chaotic cone in particle's velocity
space. Thus, the motion in this model has always a chaotic component which,
however, is never ergodic on the energy surface $H=const$. The chaotic
component is bounded by the {\it chaos border} at $\beta_0=\beta_b$. 

All particles within the chaos cone will be eventually lost diffusing to 
smaller $\beta_0$ which correspond to large amplitudes $a_x\approx\beta_0^
{-2}$. The diffusion rate in $p$ per map's iteration is obtained from
Eq.(3.13):
$$
  D_p\,=\,<(\Delta p)^2>\,\approx\,\frac{k^2}{2}\,\approx\,
  \frac{2\pi^2}{9}\,{\rm e}^{-\,\pi\lambda} \eqno (3.17)
$$
Particle's life time within the cone can be roughly estimated in the number
of $x$--oscillations as
$$
  N_x\,\sim\,\frac{P_b^2}{D_p}\,\sim\,\lambda^2\,{\rm e}^{\lambda} 
  \eqno (3.18)
$$
It is fairly long for big $\lambda\gg 1$. Besides, most particles are in
stable region $\beta_0>\beta_b\ll 1$ (3.16) and are confined there forever.
So, Budker's adiabatic magnetic trap turns out to be a very good confinement
device indeed (at least for a single particle!). 

A peculiar feature of model (3.2) is 'open' (infinite) energy surfaces
($x^2\to\infty$ if $y^2\to 0$). Moreover, ergodic (microcanonical) measure
$\Gamma_E$ of energy surface is also infinite. It is defined by the integral
$$
  \Gamma_E\,=\,\int\,\delta (H\,-\,E)\,d\Gamma \eqno (3.19)
$$
where $E$ is a particular value of energy, and $d\Gamma =dn_x\,dn_y\,
d\phi_x\,d\phi_y$ stands for the element of the full phase space. Using
$dE/dn_x=\omega_x\sim E/n_x\sim n_y\,E^{-1/2}$ and integrating over phases
and energy we obtain
$$
  \Gamma_E\,\sim\,\sqrt{E}\,\int\,\frac{dn_y}{n_y}\,=\,\sqrt{E}\,|P|\,\to\,
  \infty \eqno (3.20)
$$
which diverges as $n_y\to 0$. 

Notice that ergodic measure $\Gamma_E$ is proportional to measure $\Gamma_2$
for both maps, global (3.12) and local (3.13). This ensures a correct
description of the global dynamics by the map. If we would change dynamical
variables, e.g. $P\to n_y$, it were no longer the case, and only local map 
could be used. For this reason the special ('preferable') variables ($P,\ 
\varphi$ in our example) are called {\it ergodic variables} [24]. 

Generally, the description in discrete time (map, or difference equations)
and in continuous time (differential motion equations) is not completely
identical just because of a different time variable. An interesting and
instructive example is Lyapunov's exponent $\Lambda$ (Lecture 2) in the
model under consideration. For the standard map it depends (as anything else)
on the single parameter $K$ [23]:
$$
  \Lambda_l\,\approx\,\ln{\left(\frac{K}{2}\right)}, \qquad K\,>\,4
  \eqno (3.21)
$$
However, for the global map (3.12) local $\Lambda_l$ depends on momentum
(3.15) and must be averaged over the whole chaotic component:
$$
  \Lambda\,=\,\frac{\int_{P_b}^P\,\Lambda_l(P')\,dP'}{P}\,\to\,|P|\,\to\,
  \infty \eqno (3.22)
$$
Thus, Lyapunov's exponent for the map (per iteration) diverges as does 
the measure of the chaotic component (3.20). 

The result drastically changes in continuous time. Now, we must divide local
$\Lambda_l$ (3.21) by the half--period of $x$--oscillation $\pi /\omega_x
\sim\sqrt{E}\,\exp{(-P)}$. We have
$$
  \Lambda_t\,=\,\frac{\int_{P_b}^P\,dP'\cdot\Lambda_l(P')\cdot\omega_x(P')
  /\pi}{P}\,\to\,\frac{C}{|P|}\,\to\,0 \eqno (3.23)
$$
where $C$ is some finite constant. Thus, Lyapunov's exponent per unit
continuous time is zero ! This qualitatively different result seems to imply
violation of the main condition for chaos (Lecture 2). Resolution of the
apparent contradiction is in that for any {\it finite time}
$t\sim\exp{(|P|)}$ Lyapunov's exponent $\Lambda_t\sim 1/\ln{t}$ remains finite,
and the motion is still chaotic but, apparently, with some unusual statistical
properties. 

\subsubsection{Internal dynamics of the Yang - Mills (gauge)
fields in classical approximation}

Surprisingly, this Matinyan's problem[26] for a completely
different physical system can be also represented by Hamiltonian (3.2) with
symmetrized 'potential energy':
$$ U\,=\,\frac{(1\,+\,x^2)\,y^2\,+\,(1\,+\,y^2)\,x^2}{2} \eqno (3.24)
$$
The dynamics is always chaotic with divided phase space similar to model (3.2)
[27]. Model (3.24) describes the so--called massive gauge field that is one 
with the quanta of nonzero mass (in the classical limit!). 

The massless field corresponds to the 'potential energy'
$$ U\,=\,\frac{x^2\,y^2}{2} \eqno (3.25)
$$
and looks ergodic in numerical experiments. This model can be analysed as a
limiting case $\mu\to 0$ of Budker's model (3.2) with additional parameter
$\mu$ in the potential energy
$$
  U_{\mu}\,=\,\frac{(\mu^2\,+\,x_{\mu}^2)\,y_{\mu}^2}{2} \eqno (3.26)
$$
To this end we change variables: $x_{\mu}=\mu x,\ y_{\mu}=\mu y,\ t_{\mu}=
t/\mu$, which brings the new Hamiltonian $H_{\mu}=\mu^4\,H$ into the form of
old one (3.2), and we can use the results above. Particularly, adiabaticity
parameter $\lambda =\mu^2\,\lambda_{\mu}$ (3.3) decreases with $\mu$ for a 
given energy $H_{\mu}$. Hence, the chaos cone $\beta_b$ (3.16) rapidly expands
covering eventually the whole velocity space in agreement with numerical
experiments [26, 27].

For any finite $\mu$ the energy surfaces are also open and infinite 
in measure, and Lyapunov's
exponent takes the opposite limits in discrete and in continuous time.
This is not the case for massive field (3.24). Here the energy surfaces are
closed and finite while Lyapunov's exponent does not qualitatively depend
on the time variable albeit it may have different values in both cases
which is not important for the nature of the motion.

\subsubsection{\bf Perturbed Kepler motion} 

This is a particular case of the famous 3--body 
problem. Now we understand why it has not been solved since Newton: the
chaos is generally present in such a system. One particular example is the
motion of comet Halley perturbed by Jupiter which was found to be chaotic
with estimated life time in the Solar system of the order of 10 Myrs [28], and
with very complicated divided phase space.

The simplest model is described by a global map similar to (3.12):
$$
\begin{array}{l}
  \overline{E}\,=\,E\,+\,\epsilon\cdot F(\varphi )\\
  \overline{\varphi}\,=\,\varphi\,+\,G(\overline{E})    
\end{array} \eqno (3.27)
$$
where $E<0$ is comet's total energy, and $\varphi$ stands for Jupiter's
phase (angle) on its round orbit of unit radius (and unit velocity) at the
moment when comet is in perihelion where the perturbation effect is the
strongest. Function $G(E)=2\pi\Omega (-2E)^{-3/2}$, where $\Omega$ is
Jupiter's orbital frequency, is the Kepler law. In our units $\Omega =1$
but we will keep it in the expressions for the next example. 
Perturbation parameter
$\epsilon\approx 2\times 10^{-3}$ is essentially determined by the ratio
of Jupiter and the Sun masses ($\approx 10^{-3}$). Actually, it is somewhat 
larger because of close encounters between Jupiter and the comet depending
on the relative position of their orbits. For simplicity we assume in Eq.(3.27)
$F(\varphi )\approx\sin{\varphi}$ like in Eq.(3.12) albeit the actual
dependence is somewhat different owing to the same close encounters. 
A relatively weak perturbation by Saturn was also found to be important for
global dynamics of comet Halley.

Because of negligible comet mass the perturbation from Jupiter is fixed,
which corresponds to a time--dependent Hamiltonian, and phase $\varphi$
is simply proportional to time. In such a case the ergodic variable was
shown to be the energy ($E$, canonically conjugated to map's phase
$\varphi$) rather than comet's action in continuous time [29].

The stability parameter $K$ of the local (standard) map for model (3.27)
reads:
$$
  K\,=\,\frac{3\pi\,\epsilon\Omega}{2\sqrt{2}\,|E|^{5/2}}\,>\,1 \eqno (3.28)
$$
Chaotic component corresponds to higher energies $E>-E_b\ (|E|<E_b)$ and
goes up to $E=0$ when the comet will leave out (or was captured into) the
Solar system. In the latter case the whole motion (capture -- diffusion --
ejection) is a sort of delayed (on the diffusion stage) scattering of comet
by the Solar system.

From Eq.(3.28) the chaos border for comet Halley is roughly at $E_b\approx 
0.13$ or, in frequency, $\omega_b=(2E_b)^{3/2}\approx 0.13$. The actual 
comet frequency is
now $\omega_H\approx 0.16$ which is close to chaos border where the structure
of phase space is very complicated, with many stable domains of various size.

Detailed studies [28] have shown that current $\omega_H$ is only 5\% apart
from the border of a big stable region. Additional perturbations, including
ones of unknown nature, both in future as well as in the past could change
the character of comet's motion from chaotic to regular and vice versa.
Neglecting this possibility, the comet life time $t_H$ in the Solar system
can be roughly estimated from inhomogeneous diffusion equation (see Eq.(3.27)):
$$
  \frac{d<(\Delta E)^2>}{dt}\,\approx\,\frac{\epsilon^2}{2}\cdot
  \frac{\omega}{2\pi}\,\sim\,2E\dot{E}\,=\,\frac{\omega^{1/3}}{3}\,
  \dot{\omega} \eqno (3.29)
$$
which gives
$$
  t_H\,\sim\,4\pi\,\frac{\omega_H^{1/3}}{\epsilon^2}\,=\,
  1.7\times 10^6\,=\,3.2\,Myrs \eqno (3.30)
$$
This is essentially less than the result from computer simulation of map
(3.27): $t_H\sim 10^7\ yrs$. The difference is explained by an anomalously
slow diffusion near the chaos border.

Another example of the perturbed Kepler dynamics is a new, diffusive, 
mechanism for ionization of the Rydberg
(highly excited) Hydrogen atom in the external monochromatic electric field. 
It had been discovered in laboratory
experiments [30], and was explained by the dynamical chaos in classical
approximation [31]. In this system a given field plays a role of the 
third body. The simplest model of the diffusive photoelectric effect has
1.5 freedoms, and is described by exactly the same global map (3.27) [32],
now with $F(\varphi )=\sin{\varphi}$ but, of course, with different
perturbation parameter
$$
  \epsilon\,\approx\,\frac{2.6\,f}{\Omega^{2/3}} \eqno (3.31)
$$
where $f$ is field strength, and we use now atomic units: $|e|=m=\hbar =1$.  
Of course, this is essentially quantum problem but for a large quantum
number $n\gg 1$ (electron's action variable) the classical approximation
proved to be fairly good [31]. We will come back to the quantum effects
in this system in Lecture 4. I remind that in $n$ variable energy
$E=1/2n^2$ and Kepler frequency $\omega =1/n^3$.

Stability parameter
$$
  K\,=\,\frac{8.7\,f\,\Omega^{1/3}}{|E|^{5/2}}\,=\,50\cdot f_n\cdot 
  \Omega_n^{1/3}\,>\,1 \eqno (3.32)
$$
is expressed here in dimensionless variables $\Omega_n=\Omega\,n^3$ and
$f_n=f\,n^4$ which are reduced to the values of the corresponding atomic
quantities at energy level (action) $n$. For given field strength $f$ and
frequency $\Omega$ parameter $K$ increases with $n$. Hence, in the chaotic
component, the electron is diffusing up to eventual ionization.
If $\Omega_n\gsim 1$ the critical field $f_n\ll 1$ that is much less than
the atomic field. In the interval $1\lsim\Omega_n<n/2\gg 1$ the field
frequency may be considerably lower than that required for the conventional
(one--photon) ionization while the ionization rate is much higher provided
chaos condition (3.32).

\subsubsection{\bf Billiards and cavities}
 
In a (non--dissipative) billiard of, at least, two dimensions the ball 
motion is chaotic for almost any
shape of the boundary except special cases like circle, ellipse,
rectangle and some others (see, e.g., Refs.[2, 10]). However, 
the ergodicity (on the energy
surface) is only known for singular boundaries (of a singly--connected region).
If the boundary is smooth
enough the structure of motion becomes a very complicated admixture of
chaotic and regular domains of various size. 
In the latter case the description via global and local maps of the kind
considered above is very useful (see, e.g., Refs.[2, 33] and below). 

Another view of a billiard model
is the wave cavity in the limit of geometric
optics. This provides a helpful bridge between classical and
quantum chaos. 

Generally, the mechanism of exponential instability in billiards is related
to the particle scattering from a {\it convex} (towards the particle)
boundary [34]. A simple example is the doubly--connected region with a
convex internal boundary. The more important example is the collision of
several convex balls within any boundary which is a classical model for the
gas of molecules. 
The first simple estimate for the Lyapunov exponent in such a model was
made already by Poincare [5]:
$$
  \Lambda\,\sim\,\frac{v}{L}\,\ln{\left(\frac{L}{R}\right)}\,\leq\,
  \frac{v}{R\,{\rm e}} \eqno (3.33)
$$
where $L$ is the mean distance between the balls, and $v,\ R$ the ball
velocity and radius, respectively. The maximal instability rate is reached
at $L=R{\rm e}$.

Surprisingly, a {\it concave} boundary may also cause the instability if its
curvature is large enough [35]. This is explained by the so--called
'overfocusing': first, close trajectories converge upon reflection from 
the boundary but
later, after passing the focus, they eventually diverge. A well studied
example is the 'stadium', the planar billiard with the boundary composed
of two semicircles connected by two straight lines. For any nonzero length
of the latter the ball motion is not only chaotic but also ergodic.

Here we consider two examples of chaotic billiards with a moving
boundary. In this case the chaos is possible already in one freedom that is
for the ball motion along a straight line. One example is Ulam's model
(see Ref.[2]) for the mechanism of cosmic rays acceleration proposed by
Fermi [36]. The Fermi model was a 'gas' of huge magnetic clouds
in cosmic space and the 
protons. In the steady state the mean energy of both must be equal which
would imply an enormous acceleration of protons. Ulam checked this idea in
numerical experiments with a very simple one--freedom model: a particle
between two parallel walls, $L$ apart, one of which is oscillating with a
given velocity $V=V_0\cdot\sin{(\Omega t)}$. Surprisingly, the computation
showed no significant acceleration beyond wall's velocity $V_0$. This was
explained in Ref.[37] using the chaos theory just developed at that time.

Under condition $L\gg l=V_0/\Omega$ the particle motion is described again
by the global map (3.12) in variables $v$ (particle's velocity),
$\varphi =\Omega t$ (at collision time), and with $\epsilon =2V_0,\ G(v)
\approx 2L\Omega /v$. Hence, 
$$
  K\,\approx\,\frac{4\,L\Omega\, V_0}{v^2}\,>\,1 \eqno (3.34) 
$$
and the chaotic component is bounded from above, indeed:
$$
  \frac{v}{V_0}\,\lsim\,2\,\sqrt{\frac{L}{l}} \eqno (3.35)
$$
Acceleration $v/V_0$ turns out to be the bigger the smaller the amplitude
of the wall oscillation! It was a surprising result which would be difficult
to imagine without a theory. Of course, in the original Fermi model there was
no such restriction since the cloud motion was assumed to be random which
has been later confirmed by the chaos theory for the model of gas mentioned
above.

Dynamical variables $v,\ \varphi$ are not ergodic. Still, the local map can be
used to evaluate the conditions for chaos. If we would change the velocity
to energy the map were no longer canonical, and a more complicated map had
to be constructed. It was also assumed that the wall has infinite mass, so
that its motion is fixed. We may lift this condition to study the ergodicity
of the whole system [38]. Assume that the wall with a finite mass
$M\gg m$ (particle's mass) is a linear oscillator of frequency $\Omega$.
From the energy conservation $mv^2+MV_0^2=2E=MV_m^2$ and condition (3.34)
we can derive the chaos border $V_0=V_b$ on energy surface in the form:
$$
  \frac{V_b}{V_m}\,=\,\sqrt{1\,+\,\lambda^2}\,-\,\lambda\,, \qquad
  \lambda\,=\,2\,\frac{m}{M}\cdot\frac{L\Omega}{V_m} \eqno (3.36)
$$
where $\lambda$ may be called the ergodicity parameter. Chaotic component
corresponds to $V_0>V_b$ and increases with $\lambda$. Yet, the motion is
never completely ergodic. The measure of chaotic component can be evaluated
using the arguments applied above to Eq.(3.19). Since now the wall frequency
$\Omega =const$ is fixed $\Gamma_E\sim v$ (in continuous time). Chaos is
restricted to $v^2>v_b^2=(M/m)(V_m^2-V_b^2)$ where $v_b$ is the border value.
Hence, the relative measure of chaotic component
$$
  \Gamma_{ch}\,=\,\frac{v_b}{v_m}\,=\,\sqrt{1\,-\,\left(\frac{V_b}{V_m}
  \right)^2}\,\to\,1\,-\,\frac{1}{4\,\lambda^2} \eqno (3.37)
$$
where the latter expression corresponds to big $\lambda\gg 1$. For given 
parameters of the model the essential ergodicity is achieved in the low
energy limit only. 

Another version of Ulam's model was studied in Ref.[39] (see also Ref.[2]). 
The new model is the 'open' billiard with only a single oscillating wall in
the homogeneous field which brings the particle back to the wall. The only
difference in the global map is the phase shift between collisions:
$G(v)\approx 2v\Omega /g$ where $g$ is particle's acceleration in the field.
Then,
$$
  K\,=\,\frac{4\Omega V_0}{g} \eqno (3.38)
$$
is independent of $v$, and the chaotic acceleration becomes unbounded.
  
\subsubsection{Reversible chaos in magnetic field}

Magnetic lines can be formally considered as the 'trajectories' of some
dynamical system, the distance $s$ along a line playing a role of 'time'. 
Owing to Maxwell's equation div${\bf B}=0$ the line dynamics is Hamiltonian.
Consider a toroidal magnetic field which is used in magnetic traps, like
stellarator or tokamak, for plasma confinement [40]. 

Three--dimensional magnetic lines have $1.5$ freedoms corresponding in the 
latter example to a one--freedom oscillation (line's rotation in the plane
transverse to the torus closed axis) driven by the external perturbation
due to the variation of magnetic field in $s$ (along the axis). The transverse
surface plays here a role of the 'phase plane' for the line oscillation
which is generally nonlinear that is with the frequency depending on the
initial conditions (a distance $r$ from the axis). 

Under certain conditions the lines become chaotic [41] which is called the
'braided' magnetic field. Particularly, the lines are 'diffusing':
$$
  |\Delta r|_l\,\sim\,\sqrt{l_r\,s} \eqno (3.39)
$$
where $l_r\sim\Lambda^{-1}$ is the dynamical scale (2.16), and $\Lambda$ the
Lyapunov exponent for magnetic lines (per unit length). Notice that $s$ here 
is not restricted by the torus circumference. Instead, $s\to\infty$, and
line's diffusion is only bounded by a chaos border at large $r$, e.g.,
near the current wires producing the magnetic field. 

In sufficiently strong $B$ electron's Larmor radius $\rho$ is negligibly
small, and the electron follows a magnetic line: $s_e\approx v_{\Vert}t$
where $v_{\Vert}$ is the longitudinal velocity. Hence, the electron is also
diffusing:
$$
  |\Delta r|_e\,\sim\,\sqrt{l_r\,v_{\Vert}t} \eqno (3.40)  
$$
and it will be evetually lost. 

Now, consider the impact of electron's collisions with other particles 
in plasma [42]. For a small Larmor radius the main collision effect would be
the electron {\it velocity reversal} ($v_{\Vert}\to -v_{\Vert}$) which is
equivalent to the {\it time reversal} for magnetic lines ($s\to -s$). 
Neglecting again a finite Larmor radius, the electron will follow back the
same line. If the time reversals were periodic so would be the electron
motion as well, and the diffusion were completely stopped (in this 
approximation). However, the collisional time reversals is a random process.
Hence, the 'time' spread
$$
  |\Delta s|\,\sim\,\sqrt{l_s\,s} \eqno (3.41)
$$
where $l_s$ is the mean scattering length, would itself grow only diffusively.
This implies an anomalously slow electron diffusion (cf. Eq.(3.39)):
$$
  |\Delta r|_e\,\sim\,\sqrt{l_r\,|\Delta s|}\,\sim\,\sqrt{l_r\,\sqrt{
  l_s\,v_{\Vert}t}} \eqno (3.42)
$$

Various perturbations destroy the exact reversibility of the {\it electron}
motion. Let us consider the impact of a finite Larmor radius $\rho$. Then,
the deviation would grow exponentially up to
$$
  |\delta r|\,\sim\,\rho\cdot\exp{(\Lambda\,l_s)}\,\lsim\,l_r \eqno (3.43)
$$
at the next collision. The latter inequality is the condition for exponential,
rather than diffusive, divergence of trajectories. This is to be compared
with the collisionfree diffusion (3.40) for $v_{\Vert}t=l_s$:
$$
  \frac{(\delta r)^2}{(\Delta r)_e^2}\,\sim\,(\rho\Lambda )^2\cdot\frac{
  \exp{(2\Lambda l_s)}}{\Lambda l_s}\,\leq\,\frac{{\rm e}}{2}\,\left(\frac{
  \rho}{l_s}\right)^2\,\ll 1 \eqno (3.44)
$$
The minimum is reached at $2\Lambda l_s=1$ if $l_s\gsim\rho$ to satisfy
inequality in Eq.(3.43). 

The strong diffusion suppression is a striking manifestation of the time 
reversibility in dynamical chaos. Notice that a finite residual {\it electron}
diffusion is the result of a {\it partial reversal} of its velocity
($v_{\Vert}\to -v_{\Vert}$ only).

\subsection{Critical phenomena in dynamics}

The examples considered above suggest that a few--freedom chaotic dynamical
system has typically the divided phase space with many chaos borders.
Each of those is characterized by the so--called
{\it critical structure} [43] which is a hierarchy of chaotic
and regular domains on ever decreasing spatial and frequency scales. 
This makes statistical description a very difficult problem. Particularly,
any averaging has to be done over the chaotic component of the motion
whose measure is no longer simple Hamiltonian $\Gamma_E$ (3.19) as for
ergodic motion. Nevertheless, the critical structure can be
universally described in terms of {\it renormalization group} which 
proved to be 
so efficient in other branches of theoretical physics. In turn, such a 
renormgroup
may be considered as an abstract dynamical system which describes the variation
of the whole motion structure, for the original dynamical system, in dependence
of its spatial and temporal scale. Logarithm of the latter plays a role of
'time' (renormtime) in that renormdynamics. At the chaos border the latter is
determined by the motion frequencies. The simplest renormdynamics is a periodic
variation of the structure or, for a renorm--map, the invariance of the
structure with respect to the scale [44]. Surprisingly, this scale invariance
includes the chaotic trajectories as well. The opposite limit --
{\it renormchaos} -- 
is also possible, and was found in several models (see Ref.[43]).
Remarkably, for a two--dimensional map, which also may describe the
two--freedom conservative system, an extremely complicated renormdynamics
can be reduced to a most simple one--dimensional map
$$
  \overline{r}\,=\,\frac{1}{r}\ \ \ mod\,1 \eqno (3.45)
$$
where $r$ is the so--called rotation number that is the ratio of the two
motion frequencies [43]. This map was introduced by Gauss in the number theory
and has been well studied by now [10]. Particularly, the Lyapunov exponent
(per iteration) $\Lambda =\pi^2/6\ln{2}$, and almost any intial $r_0$
generates a random trajectory which corresponds to random fluctuations of the
motion structure from one scale to the next. Exceptional rationals $r_0=m/n$
give rise to a periodic oscillation of the structure and, hence, to scale
invariance in $n$ steps. 

Even though the critical structure occupies a very narrow strip along the 
chaos border it may qualitatively change the statistical properties of the
whole chaotic component. This is because a chaotic trajectory unavoidably
enters from time to time the critical region and 'sticks' there for a time
the longer the closer it comes to the chaos border. The sticking results in
a slow power--law, rather than exponential, correlation decay for large time:
$$
  C(\tau)\,\sim\,\tau^{-\,p_C}\,, \quad \tau\,\to\,\infty \eqno (3.46)
$$
Moreover, exponent $p_C<1$, and for the two--dimensional map was found
numerically to be approximately $p_C\approx 0.5$ in agreement with a
simple theoretical analysis [43]. In higher dimension the dependence
$$
  p_C\,=\,\frac{1}{2N\,-\,2} \eqno (3.47)
$$
was conjectured based on the same physical theory. Here $N$ is the number of
linearly independent (incommensurate) frequencies both internal (unperturbed)
and driving. 

Slow decaying correlation (3.46) implies a singular power spectrum which is
the Fourier transform of $C(\tau )$:
$$
  S(\omega )\,\sim\,\frac{1}{\omega^{p_S}}\,,\ \omega\,\to\,0\,;
  \qquad p_S\,=\,1\,-\,p_C\,=\,\frac{2N\,-\,3}{2N\,-\,2} \eqno (3.48)
$$
As $N\to\infty$ the spectrum approaches that of 'mysterious' $1/\omega$ noise
(see, e.g., Ref.[45]). In the minimal dimension ($N=2$) the singular 
spectrum is $S\sim 1/\sqrt{\omega}$.

The diffusion determined by correlation (3.46) turns out to be anomalously 
fast [46] as the standard diffusion rate
$$
  D\,\sim\,\int\,C(\tau )\,d\tau\,\to\,\infty \eqno (3.49)
$$
diverges for $p_C\leq 1$. In such a case the dispersion $\sigma^2$ (second
moment of the distribution function, e.g., $\sigma^2=<(\Delta P)^2>$ in the
example below) is given by a double integral of correlation or by the
differential equation
$$
  \frac{d^2\,\sigma^2}{d\tau^2}\,=\,2\,C(\tau ) \eqno (3.50)
$$
which can be applied to a more general problem [47]. A particular example is
the standard map (3.13) (in variables $P=Tp,\ \varphi$) for special values
of parameter $K=K_n\approx 2\pi n$ with any integer $n\geq 1$ [23].  
In this case there are two fixed points $\varphi =\varphi_1=const$ satisfying
$K\cdot\sin{\varphi_1}=\pm 2\pi n$, and momentum $|P|$ growing proportionally
to (discrete) time. The fixed points are stable but relative area of both
stable domains around is rather small: $A_n\approx 8/\pi^2K_n\approx 
2/\pi^4n^2$ and rapidly decreases with $n$. The biggest one is $A_1\approx 
2\%$ only. Within a stable region the particle is accelerating independent
of initial conditions. This is how the so--called microtron works, the first
cyclic accelerator for relativistic electrons proposed by Veksler [48].

More interesting is the behavior of a chaotic trajectory. From time to time it
approaches the chaos border of a tiny stable domain and sticks there for 
a while being accelerated much more rapidly than in the rest ($98\%$!) of the
chaotic component. Since there are two stable domains with opposite 
acceleration the resulting motion would be also diffusive but anomalously fast:
for $p_C=1/2$ the average increase in momentum becomes 
$|\Delta P|\sim t^{p_D}$ with
$p_D=2-p_C=3/2$ [46]. A more accurate calculation leads to the relation [25]:
$$
  <(\Delta P)^2>\,\approx\,\frac{\alpha}{2}\,A_n\,K_n^2\,t^{3/2}\,\approx\,
  \frac{4\alpha}{\pi^2}\,t^{3/2} \eqno (3.51)
$$
Here $t$ is map's discrete time, and $\alpha\approx 0.5$ is taken from
numerical experiments [49] where such enhanced diffusion was observed for the
first time. Actually, the normal diffusion rate $D=<(\Delta P)^2>/t$ 
was measured and found to be $100$ times (!) larger than expected
$D=K^2/2$. Remarkably, the rate of anomalous diffusion (3.51) does not
depend on stable area $A_n$. Yet, the crossover time 
$t_a\approx\pi^8n^4\sim A_n^{-2}$
from normal to anomalous diffusion does so.

In higher dimension $p_D(N)=(4N-5)/(2N-2)\to 2$ for $N\gg 1$, and
$|\Delta P|\sim t$. This is the fastest homogeneous diffusion possible.
The motion would be close to the straight acceleration but in both 
directions of $P$ variation!

\section{Quantum pseudochaos}

The mathematical theory of dynamical chaos -- ergodic theory -- is 
selfconsistent. However, this is not the case for the physical theory unless
we accept the philosophy of the two separate mechanics, classical and quantum.
Even though such a view cannot be excluded at the moment it has a profound 
difficulty concerning the border between the two. Nor is it necessary
according to recent intensive studies of quantum dynamics. Then, we have 
to understand the mechanics of dynamical chaos from the quantum point of view.
Our guiding star will be the {\it correspondence principle} which requires
the complete quantum theory for any classical phenomenon, in the quasiclassical
limit, assuming that the whole classical mechanics is but a special part
(the limiting case) of currently most general and fundamental physical theory,
the quantum mechanics. Now it would be more correct to speak about the
quantum field theory but here I restrict myself to finite--dimensional
systems only.

\subsection{The correspondence principle}

In attempts to build up the quantum theory of dynamical chaos we immediately
encounter a number of apparently very deep contradictions between the well
established properties of classical dynamical chaos and the most fundamental
principles of quantum mechanics.

To begin with, the quantum mechanics is commonly understood as a 
{\it fundamentally statistical theory} which seems to imply {\it always}
some quantum chaos, independent of the behavior in the classical limit. 
This is
certainly true but in some restricted sense only. A novel development here
is the {\it isolation} of this fundamental quantum randomness as solely
the characteristic of a very specific quantum process, the measurement, and
even as the particular part of that - the so--called {\it $\psi$--collapse}
which, indeed, has so far no dynamical description.

No doubt, the quantum measurement is absolutely necessary for the study of
microworld by us, the macroscopic human beings. Yet, the measurement is, in a
sense, foreign to the proper microworld which might (and should) be described
separately from the former. Explicitly[4] or, more often, implicitly such a
philosophy has become by now common in the studies of chaos but not yet
beyond this field of research (see, e.g., Ref.[50]).

This approach allows us to single out the dynamical part of quantum mechanics
as represented by a {\it specific dynamical variable} $\psi (t)$ in the
{\it Hilbert space} satisfying some {\it deterministic} equation of motion,
e.g., the Schr\"odinger equation. The more difficult and vague statistical
part is left behind for a better time. Thus, we temporarily bypass 
(not resolve!)
the first serious difficulty in the theory of quantum chaos.
The separation of the first part of quantum dynamics, 
which is very natural from mathematical viewpoint, had been first introduced 
and emphasized by Schr\"odinger who, however, certainly underestimated 
the importance of the second part in physics.

However, another principal difficulty arises. As is well known, the energy
(and frequency) spectrum of any  quantum motion {\it bounded in phase space} 
is always {\it discrete}. And this is not the property of a particular equation
but rather a consequence of the fundamental quantum principle - the 
{\it discreteness of phase space} itself, or in a more formal language, the
noncommutative geometry of quantum phase space. Indeed, according to another
fundamental quantum principle -- the uncertainty principle -- a single quantum
state cannot occupy the phase space volume $V_1\lsim\hbar^N\equiv 1$ (in what
follows I set $\hbar =1$).
Hence, the motion bounded in a domain of volume $V$ is represented
by $V/V_1\sim V$ eigenstates, the property even stronger than the general
discrete spectrum (almost periodic motion). 

According to the existing ergodic theory such a motion is considered to be
{\it regular} which is something opposite to the known chaotic motion with
continuous spectrum and exponential instability, again 
independent of the classical behavior. This seems to {\it never} imply any
chaos or, to be more precise, {\it any classical--like chaos} as defined 
in the ergodic theory. Meanwhile, the correspondence principle requires
{\it conditional chaos} related to the nature of motion in the classical 
limit. 

\subsection{Pseudochaos}

Now the principal question to be answered reads: where is the expected quantum
chaos in the ergodic theory? Our answer to this question[51] (not commonly
accepted as yet) was concluded from a simple observation (principally well 
known but never comprehended enough) that the sharp border between the
discrete and continuous spectrum is
physically meaningful in the limit $|t|\to\infty$ only, the condition
actually assumed in the ergodic theory. Hence, to understand the quantum chaos
the existing ergodic theory needs some modification by introducing a new
'dimension', the time. In other words, a new and central problem in the ergodic
theory becomes the {\it finite--time statistical properties} of a dynamical 
system, both quantum as well as classical.

Within a finite time the discrete spectrum is dynamically equivalent to the
continuous one, thus providing much stronger statistical properties of the
motion than it was (and still is) expected in the ergodic theory in case of
discrete spectrum. In short, the motion with discrete spectrum may exhibit
{\it all} the statistical properties of the classical chaos but only on
some {\it finite} time scales. 

A simple example of (classical) pseudochaos is a symbolic trajectory 
(Lecture 2) of some period $T_s$ composed of {\it random} elements $m_i\ 
(i=1,...,T_s)$ whatever the origin of the randomness. In any event, {\it most}
finite sequences $m_i$ are {\it random}, indeed, according to the algorithmic
theory of dynamical systems [21]. In this example there is a {\it single}
time scale ($T_s$) for {\it all} statistical properties while, generally, 
there are several {\it different} scales related to a particular property
(see below).

The conception of time scale is a
fundamental one in our theory of quantum chaos [51]. This is certainly
a {\it new} dynamical phenomenon, related but not identical at all to the
classical dynamical chaos.\footnote{There are very special, even exotic I
would say, examples of the 'true', classical--like, chaos in quantum systems
(see [51, 4] and references therein). In all such cases the quantum motion
is not only unbounded in some phase space variables but, moreover, the
latter grow exponentially in time.}
 We call it {\it pseudochaos}, the term {\it pseudo}
intending to emphasize the difference from the asymptotic (in time) chaos
in the ergodic theory. Yet, from the physical point of view, we accept here,
the latter, strictly speaking, does not exist in the Nature. So, in the
common philosophy of the universal quantum mechanics the {\it pseudochaos
is the only true dynamical chaos} (cf. the term pseudoeuclidian geometry
in special relativity). The asymptotic chaos is but a limiting pattern which
is, nevertheless, very important both in the theory to compare 
with the real chaos
and in applications as a very good approximation in macroscopic domain as is
the whole classical mechanics.
Ford calls it {\it mathematical chaos} as contrasted to the {\it real
physical chaos} in quantum mechanics [52]. Another curious but impressive
term is {\it artificial reality} [53] which is, of course, a selfcontradictory
notion reflecting, particularly, confusion in the interpreting such surprising
phenomena like chaos.

Until recently the conception of classical dynamical chaos was completely
incomprehensible, especially for physicists. One particular point of
confusion was (and still remains to some extent) the Second Law of
thermodynamics, the entropy increase in a closed system. Meanwhile,
the entropy defined by the {\it exact} phase density is the motion integral
in any Hamiltonian system. Some physicists are still reluctant to assume
thermodynamic entropy determined via the {\it coarse--grained} density in which
case it may well increase under conditions of dynamical chaos.
From many researchers I know that they actually observed dynamical chaos
in numerical or laboratory experiments but... did their best to get rid of it
as some artifact, noise or other interference! Now the situation in this field
is upside down: most researchers (not me!) insist that if an apparent chaos
is not like that in the classical mechanics (and in the existing 
ergodic theory)
then it is not a chaos at all. The most controversial conception in today's
disputes is just the quantum chaos. The curiosity of the current 
situation is that
in most studies of the 'true' (classical) chaos the digital computer is used
where only {\it pseudochaos} is possible that is one like in {\it quantum}
(not classical) mechanics!

The statistical properties of the discrete--spectrum motion is not a 
completely new subject of research, it goes back to the time of intensive
studies in the mathematical foundations of statistical mechanics {\it before}
the dynamical chaos was discovered or, better to say, was understood (see, 
e.g., Ref.[54]). We call this early stage of the theory {\it traditional
statistical mechanics} (TSM). It is equally applicable to both classical as
well as quantum systems.
For the problem under consideration here one of the most
important rigorous results with far--reaching implications was the
{\it statistical independence} of oscillations with incommensurate 
(linearly independent) frequencies $\omega_n$, such that the only solution of
the resonance equation 
$$ \sum_n^N\,m_n\cdot\omega_n\,=\,0 \eqno (4.1)
$$
in integers is $m_n\equiv 0$ for all $n$. This is a generic property of the 
real numbers that is the resonant frequencies (4.1) form a set of zero Lebesgue
measure. If we define now $y_n=\cos{(\omega_n t)}$ the statistical independence
of $y_n$ means that trajectory $y_n(t)$ is ergodic in $N$--cube $|y_n|\leq 1$.
This is a consequence of ergodicity of the phase trajectory $\phi_n (t)=
\omega_n t\ mod\ 2\pi $ in $N$--cube $|\phi_n|\leq \pi$. 

Statistical independence is a basic property of a set to which the probability
theory is to be applied. Particularly, the sum of statistically independent
quantities
$$ x(t)\,=\,\sum_n^N\,A_n\cdot\cos{(\omega_n\,t\,+\,\phi_n)} \eqno (4.2)
$$
which is the motion with discrete spectrum, is a typical object of this theory.
However, the familiar statistical properties like Gaussian fluctuations,
postulated (directly or indirectly) in TSM, are reached in the limit
$N\to\infty$ only[54] which is called {\it thermodynamic limit}. In TSM this
limit corresponds to infinite--dimensional models [10] which provide a
very good approximation for macroscopic systems, both classical and quantal.

However, what is really necessary for good statistical properties of sum (4.2)
is a big number of frequencies $N_{\omega}\to\infty$ which makes the discrete
spectrum continuous (in the limit). In TSM the latter condition is satisfied
by setting $N_{\omega}=N$. The same holds true for quantum fields which are
infinite--dimensional.
In quantum mechanics another mechanism, independent
of $N$, works in the quasiclassical region $q\gg 1$ where $q=n/\hbar\equiv n$
is some big quantum parameter, e.g. quantum number, and $n$ stands for a
characteristic action of the system. Indeed, if the quantum motion (4.2)
(with $\psi (t)$ instead of $x(t)$) is determined by many ($\sim q$) 
eigenstates we can set $N_{\omega}=q$ independent of $N$. The actual number of 
terms in expansion (4.2) depends, of course, on a particular state $\psi (t)$
under consideration. For example, if it is just an eigenstate the sum reduces
to a single term. This corresponds to the special peculiar trajectories of
classical chaotic motion whose total measure is {\it zero}. Similarly, in
quantum mechanics $N_{\omega}\sim q$ for {\it most} states if the system is
{\it classically chaotic}. This important condition was found to be certainly
{\it sufficient} for good quantum statistical properties (see Ref.[51] and
below). Whether it is also a necessary condition remains as yet
unclear.

Thus, with respect to the mechanism of the quantum chaos we essentially
{\it come back} to TSM with exchange of the number of freedoms $N$ for 
quantum parameter $q$. However, in quantum mechanics we are not interested,
unlike TSM, in the limit $q\to\infty$ which is simply the classical mechanics.
Here, the central problem is the statistical properties for {\it large but
finite} $q$. This problem does not exist in TSM describing macroscopic systems.
Thus, with an old mechanism the new phenomena were understood in quantum
mechanics.

The direct relation between these two seemingly different mechanisms 
of chaos can be traced 
back in some specific dynamical models [4]. One interesting example is the 
nonlinear Schr\"odinger equation [88].
From a physical point of view it describes the motion of a quantum system 
interacting with many other freedoms whose state is expressed via the 
$\psi$ function of the system itself 
(the so--called mean field approximation). This approximation becomes exact 
in the limit $N\to\infty$ which is 
a particular case of the thermodynamic limit. Therefore, the 
mechanism for chaos in this system is apparently
the old one. On the other hand, the nonlinear Schr\"odinger equation 
has generally exponentially 
unstable solutions, hence the mechanism of chaos here seems to be the new one. 
Thus, for this particular 
model both mechanisms describe the same physical process. 
 We would like to emphasize that the 'true' chaos present in 
these apparently few--dimensional models actually refers 
to infinite--dimensional systems.

\subsection{Characteristic time scales in quantum chaos}

The existing ergodic theory is asymptotic in time, and hence contains no time 
scales at all. There are two reasons for this. One is technical: it is much 
simpler to derive the asymptotic relations than to obtain rigorous 
finite--time estimates. Another reason is more profound. All statements in
the ergodic theory hold true up to measure zero that is excluding some peculiar
nongeneric sets of zero measure. Even this minimal imperfection of the theory
did not seem completely satisfactory but has been 'swallowed' eventually and
is now commonly tolerated even among mathematicians to say nothing about
physicists. In a finite--time theory all these exceptions acquire a {\it small
but finite} measure which would be already 'unbearable' (for mathematicians).
Yet, there is a standard mathematical trick, to be discussed below, for
avoiding both these difficulties. 

The most important time scale $t_R$ in quantum chaos is given by the general
estimate
$$ \ln{t_R}\,\sim\,\ln{q}\,, \qquad t_R\,\sim\,q^{\alpha}\,\sim\,\rho_0\,
   \leq\,\rho_H \eqno (4.3)
$$
where $\alpha\sim 1$ is a system--dependent parameter. This is called the
{\it relaxation time scale} referring to one of the principal properties of
the chaos -- {\it statistical relaxation} to some steady state (statistical
equilibrium). The physical meaning of this scale is principally simple, and
it is directly related to the fundamental uncertainty principle ($\Delta t
\cdot\Delta E\,\sim\,1$) as implemented in the second Eq.(4.3) where $\rho_H$
is the {\it full} average energy level density (also called Heisenberg time).
For $t\lsim t_R$ the discrete spectrum is not resolved, and the statistical
relaxation follows the classical (limiting) behavior. This is just 
the 'gap' in the
ergodic theory (supplemented with the additional, time, dimension) where
the pseudochaos, particularly quantum chaos, dwells. A more accurate estimate
relates $t_R$ to a {\it part} $\rho_0$ of the level density. This is the
density of the so--called {\it operative eigenstates} only that is those
which are actually present in a particular quantum state $\psi$, and which
actually control its dynamics. 

The formal trick mentioned above is to consider not finite--time relations we 
really need in physics but rather the special {\it conditional limit} 
(cf. Eq.(2.18)):
$$ t,\,q\,\to\,\infty\,, \qquad \tau_R\,=\,\frac{t}{t_R(q)}\,=\,const 
   \eqno (4.4)
$$
Quantity $\tau_R$ is here a new rescaled time which is, of course, nonphysical
but very helpful technically. The {\it double} limit (4.4) (unlike the single
one $q\to\infty$) is {\it not} the classical mechanics which holds true, in
this representation, for $\tau_R\lsim 1$ and with respect to the statistical
relaxation only. For $\tau_R\gsim 1$ the behavior becomes essentially quantum
(even in the limit $q\to\infty$ !) and is called nowadays {\it mesoscopic
phenomena}. Particularly, the quantum steady state is quite different from
the classical statistical equilibrium in that the former may be {\it localized}
(under certain conditions) that is {\it nonergodic} in spite of classical
ergodicity. 

Another important difference is in {\it fluctuations} which are
also a characteristic property of chaotic behavior. In comparison with 
classical mechanics the quantum $\psi (t)$ plays, in this respect, 
an intermediate
role between the classical trajectory (exact or symbolic) with big relative
fluctuations $\sim 1$ and the coarse--grained classical phase space density
with no fluctuations at all. Unlike both the fluctuations of $\psi (t)$ are
$\sim N_{\omega}^{-1/2}$ which is another manifestation of statistical 
independence, or {\it decoherence}, of even pure quantum state (4.2) in case
of quantum chaos. In other words, chaotic $\psi (t)$ represents statistically
a {\it finite ensemble} of $\sim N_{\omega}$ systems even though formally
$\psi (t)$ describes a single system. Quantum fluctuations clearly demonstrate 
also the difference between physical time $t$ and auxiliary variable $\tau$:
in the double limit ($t,\,q\to\infty$) the fluctuations vanish, and one needs
a new trick to recover them.

The popular term {\it mesoscopic} means here an {\it intermediate} behavior
between classical ($q\to\infty$) and quantum (e.g., localization) one.
In other words, in a mesoscopic phenomenon both classical and quantum
features are combined simulteneously. Again, the correspondence principle
requires transition to the completely classical behavior. This is, indeed, 
the case according to Shnirelman's theorem or, better to say, to a
physical generalization of the theorem [55]. Namely, the mesoscopic
phenomena occur in the so--called {\it intermediate quasiclassical asymptotics}
where $q\gg 1$ is already very big but still $q\lsim q_f$ less than a certain
critical $q_f$ which determines the border of transition to a fully classical
behavior. The latter region, ensured by the above theorems in accordance with
the correspondence principle, is called the {\it far quasiclassical 
asymptotics}. 

The striking well known examples of mesoscopic phenomena are superconductivity
and superfluidity. A mesoscopic parameter here is the temperature which
determines the behavior of microparticles, electrons and atoms, respectively.
The far asymptotics corresponds here to $T>T_f$ where both essentially
quantum phenomena disappear. 

The relaxation time scale should not be confused with the {\it Poincare
recurrence time} $t_P\gg t_R$ which is typically much longer, and which
sharply increases with decreasing of the recurrence domain. Time scale
$t_P$ characterizes big fluctuations (for both the classical trajectory,
but not the phase space density, and the quantum $\psi$) of which recurrences
is a particular case. Unlike this $t_R$ characterizes the average relaxation
process.

More strong statistical properties than relaxation and fluctuations are
related in the ergodic theory to the exponential instability of motion. Their
importance for the statistical mechanics is not completely clear. Nevertheless,
in accordance with the correspondence principle, those stronger properties
are also present in quantum chaos as well but on a {\it much shorter} time
scale
$$ t_r\,\sim\,\frac{\ln{q}}{h} \eqno (4.5)
$$
where $h$ is classical metric entropy (2.11). This time scale was discovered 
and partly explained in Refs.[57] (see also Refs.[51, 4]). 
We call it {\it random time scale}.
Indeed, according to the Ehrenfest theorem the motion of a narrow wave packet
follows the beam of classical trajectories as long as the packet remains
narrow, and hence it is as random as in the classical limit. 
Even though the random time scale is very short, it grows indefinitely as 
$q\to\infty$. Thus, a
temporary, finite--time quantum pseudochaos turns into the classical dynamical 
chaos in accordance with the correspondence principle. 
Again, we may consider the {\it conditional limit}:
$$ t,\,q\,\to\,\infty\,, \qquad \tau_r\,=\,\frac{t}{t_r(q)}\,=\,
   const  \eqno (4.6)
$$
Notice that {\it scaled} time $\tau_r$ is different from $\tau_R$ in Eq.(4.4).

Particularly, if we fix time $t$, then in the limit $q\to\infty$ we obtain 
the transition to the classical instability in accordance with 
the correspondence 
principle while for $q$ fixed, and $t\to\infty$ we have the proper quantum 
evolution in time. For example, the quantum Lyapunov exponent
$$ 
   \Lambda_q(\tau_r)\,\to\,\left\{\begin{array}{ll}
   \Lambda\, , \quad & \tau_r\,\ll\, 1 \\
   0\,,        \quad & \tau_r\,\gg\,1 \end{array} \right.  \eqno (4.7)
$$
 
The quantum instability ($\Lambda_q>0$) was observed in numerical experiments,
indeed [58, 4]. What does terminate the instability for $t\gsim t_r$? 
A naive explanation that the major size of the originally most narrow
quantum packet reaches the full swing of a bounded motion is obviously too
simplified. This is immediately clear from the comparison with the classical
packet behavior (Lecture 2). Also, the quantum packet squeezing is not
principally restricted since only 2--dimensional area (per freedom) is
bounded from below in quantum mechanics. Instead, numerical experiments 
show that the original 
wave packet, after a considerable stretching similar to the classical one,
is rapidly destroyed. Namely, it gets split into many new small packets.
A possible explanation[59] (see also Ref.[4]) is related to the 
discreteness of the action 
variable in quantum mechanics which leads to the ``rupture'' of a very long
stretched packet into many pieces. Such a mechanism determines a new
{\it destruction time scale} which, for the quantized standard map
 (see below), is given by the estimate:
$$ t_d\,\sim\, \frac{|\ln{T}|}{2\Lambda} \eqno (4.8)
$$
This roughly agrees with the results of numerical experiments [58, 4].
 As expected $t_d\sim t_r$ (see Eq.(4.5)).
 
There is another mechanism which produces deviation of the quantum packet 
evolution
from the classical motion [59]. We call it {\it inflation} because of
the increase in time of the phase space area occupied by the quantum
phase space density (the Wigner function )
contrary to the classical density which is conserved (Liouville's theorem). 
The inflation can be analyzed using the quantum Liouville equation 
for the Wigner function $W$ [60].
In case of standard map this equation reduces to:
$$ \frac{dW(n,\,\varphi )}{dt}\,\approx\,-\,\frac{1}{24}\,\frac{\partial^3 H}
   {\partial\varphi^3}
   \,\frac{\partial^3W}{\partial n^3} \eqno (4.9)
$$
and gives the following estimate for the {\it inflation time scale}:
$$ t_{if}\,\sim\,\frac{|\ln{(TK^2/\Lambda^2)}|}{6\Lambda}   \eqno (4.10)
$$
The inflation time is of the order of destruction time (4.8) and of 
the random time scale (4.5) as well which implies, particularly, a
considerable squeezing of a wave packet.
 
An important implication of the above picture of packet's time evolution is
the rapid and complete destruction of the so--called generalized coherent
states [61] in quantum chaos.

In quasiclassical region ($q\gg 1$) scale $t_r\ll t_R$ (4.3). This leads to an
interesting conclusion that the quantum diffusion and relaxation are
{\it dynamically stable} contrary to the classical behavior. It suggests, 
in turn, that the motion instability is not important {\it during}
statistical relaxation. However, the {\it foregoing} correlation decay on
short random time scale $t_r$ is crucial for the statistical properties
of quantum dynamics.
Dynamical stability of quantum diffusion has been proved
in striking numerical experiments with time
 reversal [65]. In a classical chaotic system the diffusion is immediately
 recovered due to numerical "errors" (not random !) amplified by the local
 instability. On the contrary, the quantum "antidiffusion" proceeds untill the
 system passes, to a very high accuracy, the initial state, 
 and only than the normal
 diffusion is restored. The stability of quantun chaos on relaxation
 time scale is comprehensible as the random time scale is much shorter.
 Yet, the accuracy of the reversal (up to $\sim 10^{-15}$ (!) )
 is surprising. Apparently, this is explained
 by a relatively large size of the quantum wave packet as compared to the 
 unavoidable rounding-off errors unlike the classical computer trajectory
 which is just of that size [68]. In the standard map the size
of the optimal, least-spreading, wave packet $\Delta\varphi\sim \sqrt{T}$ [51].
 On the other hand, any quantity in the computer must well exceed 
 the rounding--off error $\delta\ll 1$. Particularly, $T\gg\delta$, and 
 $(\Delta\varphi)^2/\delta^2\gsim(T/\delta)\delta^{-1}\gg 1$.

\subsection{Quantum localization: the kicked rotator model}

The standard map (3.13) was shown in Lecture 3 to provide the local 
description of motion for many more realistic classical models. So, the 
quantized standard map seems to be a good approach in the studies of quantum
chaos as well. This can be done in two ways. The first one is to derive exact
 unitary operator  
 $\hat{U}_T$ over some time interval $T$:  
 $$ \overline{\psi (t)}\,\equiv\,\psi (t\,+\,T)\,=\,\hat{U}_T\,\psi (t),   
    \quad \hat{U}_T\,=\,  
    \exp{\left(\,-\,i\,\int^{T} dt\,\hat{H}\right)} \eqno (4.11)  
 $$  
 where $\hat{H}$ is the Hamiltonian operator.  
Generally, this is a very difficult mathematical problem which we will not
discussed (see Ref.[62]). Instead, we consider here the second way:
the direct quantization of the classical standard map (3.13) which is,
of course, only approximate solution of the whole problem. I am not aware
of any thorough analysis of the accuracy and limitations of this simple
method. However, the direct comparison of such a quantum map with the numerical
solution of Schr\"odinger equation for the diffusive photoeffect in Rydberg
Hydrogen atom confirms that the former is a reasonable approximation, 
indeed [32]. 

Quantization of standard map with Hamiltonian
$$
  H(n,\,\varphi ,\,t)\,=\,\frac{n^2}{2}\,+\,k\cdot\cos{\varphi}\cdot
  \delta_T(t) \eqno (4.12)
$$
leads to the unitary operator [63]:
$$ \hat{U}_T\,=\,\exp{\left(\,-\,i\,\frac{T\,\hat{n}^2}{2}\right)}\,\cdot\,
   \exp{(\,-\,ik\cdot\cos{\hat{\varphi}})} \eqno (4.13)
$$
where $\delta_T(t)$ is $\delta$--function of period $T$, and
$\hat{n} = - i \frac{\partial}{\partial \varphi}$.

Standard map (4.13) is defined on a cylinder ($-\infty <n<+\infty$) where
the motion can be unbounded. To describe a bounded motion in a conservative
system it is more convenient to make use of another version of the standard 
map, namely, one
on a torus with {\it finite} number of states $L\gg 1$. In momentum 
representation
$\psi (n,\,t)$ it is described by a finite unitary matrix $U_{nm}$:
$$ \overline{\psi (n)}\,=\,\sum_{m=-L_1}^{L_1} U_{nm}\,\psi (m) \eqno (4.14)
$$
where $L=2L_1+1\approx 2L_1$, and
$$ U_{nm}=\frac{1}{L}\exp{\left(i\frac{T}{4}(n^2+m^2)\right)}
   \cdot \sum_{j=-L_1}^{L_1}\exp{\left[-ik\cdot\cos{(2\pi j/L)}-
   2\pi i(n-m)j/L\right]} \eqno (4.15)
$$
while $T/4\pi =M/2L$ is now rational [64]. 
 
There are three quantum parameters in this model: perturbation $k$, period $T$
and size $L$ in momentum, but only two classical combinations remain: 
perturbation $K=k\cdot T$
and classical size $M=TL/2\pi$ which is the number of resonances over the
torus. Notice that the quantum dynamics is generally more rich than the
classical one as the former depends on an extra parameter. It is, of course,
another representation of Planck's constant which we have set $\hbar =1$.
This is why in quantized standard map we need both parameters, $k$ and $T$,
separately and cannot combine them in a single classical parameter $K$.
 
The quasiclassical region, where we expect quantum chaos, corresponds to 
$T\to 0,\ k\to\infty ,\ L\to\infty$ while the classical parameters 
$K=const$ and $M=const$.

A technical difficulty in evaluating $t_R$ for a particular dynamical problem 
is in that the density $\rho_0$ depends, in turn, on the dynamics. So, we
have to solve a self--consistent problem. For the standard map the answer
is known (see Ref.[4]):
$$ t_R\,=\,\rho_0\,=\,2D_0 \eqno (4.16)
$$
where $D_0 = k^2/2$ is the classical diffusion rate (for $K\gg 1$). 
The quantum diffusion rate depends on the scaled variable 
$$\tau_R\,=\,\frac{t}{2\,D_0(k)}
$$ 
and is given by 
$$
  D_q\,=\,\frac{D_0}{1\,+\,\tau_R}\,\to\,
  \left\{\begin{array}{ll}
  D_0\, , \quad & \tau_R\,=\,t/t_R\,\ll\, 1 \\
   0\,,        \quad & \tau_R\,\gg\,1 \end{array} \right.  \eqno (4.17)
$$
This is an example of scaling in discrete spectrum which stops
eventually the quantum diffusion. 

A simple estimate for $t_R$ in the standard map can be derived as follows [51]
(see also Ref.[67]).
The quantum map as a time--dependent system is characterised by
{\it quasienergies} which are determined modulus $\Omega =2\pi$ where $\Omega$
is the frequency of external perturbation, and where the latter value 
corresponds to the discrete time with one map's iteration as the time unit.
Then, the mean density of operative eigenstates $\rho_0=N_0/\Omega$ where $N_0$
is the number of the latter. In turn, $N_0\sim 2\sqrt{D_0t_R}$ which is also
the number of unperturbed states ($n$) covered by the quantum diffusion until
it stops. Here we assume that both quantum diffusion as well as the
eigenstates are statistically homogeneous that is they couple {\it all}
unperturbed states, at least, mesoscopically. This natural assumption is in
agreement with all the numerical experiments. Microscopic deviations from
homogeneity, the so--called 'scars' and some others (see, e.g., Refs.[69]),
apparently do not affect the mesoscopic quantum properties. Then, we arrive
at a simple estimate:
$t_R\sim D_0$ (cf. Eq.(4.16)). Moreover, the same estimate
gives also the size, or {\it localization length}, of the localized steady
state ($l_s$) as well as that of the eigenfunctions ($l$): $l_s\sim l\sim 
t_R\sim D_0$. These are remarkable relations in that they connect essentially
{\it quantum} characteristics ($l_s,\ l,\ t_R$) with the {\it classical}
diffusion rate $D_0$. This is just a characteristic feature of the mesoscopic
phenomena. 

For the standard map on the cylinder the quantum diffusion is always localized,
the shape of the localized states being approximately exponential (see, e.g.,
Ref.[4]):
$$
  \psi (n)\,\approx\,\frac{\exp{(\,-\,\frac{|n\,-\,n_0|}{l})}}
  {\sqrt{l}} \eqno (4.18)
$$
and the same for the steady state. Interestingly, two localization lengths are
different [51]:
$$
  l_s\,\approx\,D_0 \quad {\rm while} \quad l\,\approx\,\frac{D_0}{2}
  \eqno (4.19)
$$
because of big fluctuations.
  
Generally, the quantum localization is a non--universal but very 
interesting and important mesoscopic phenomenon because it means the 
formation of non--ergodic 
or localized states (both a steady state as well as eigenstates) for 
classically ergodic motion. Moreover, the localized steady state 
{\it depends} on the initial state from which the diffusion starts.
For the standard map on torus 
the {\it ergodicity parameter} controlling localization can be defined as:
$$ \lambda\,=\,\frac{D_0}{L}\,\sim\,\left(\frac{t_R}{t_e}\right)^{1/2}
   \,\sim\,\frac{k^2}{L}\,\sim\,\frac{K}{M}\cdot k \eqno (4.20)
$$
where $t_e\sim L^2/D_0$ is a characteristic time of the classical 
relaxation to the ergodic steady state $|\psi (n)|^2\approx const$.
 
If $\lambda\gg 1$ the final steady state as well as all the eigenfunctions
are ergodic that is the corresponding Wigner functions
are close to the classical microcanonical distribution in phase space (3.19).
This is far quasiclassical asymptotics. 
It can be reached, particularly, if the classical parameter
$K/M$ is kept fixed while the quantum parameter
$k\to\infty$.
 
However, if $\lambda\ll 1$ all the eigenstates and the steady state are
non--ergodic. It means that their structure remains 
essentially quantum,
no matter how large is the quantum parameter $k\to\infty$. This is
intermediate quasiclassical asymptotics or mesoscopic domain.
Particularly, it corresponds to $K>1$ fixed, $k\to\infty$ and 
$M\to\infty$ while $\lambda\ll 1$ remains small.
 
In terms of localization length the region of mesoscopic phenomena
is defined by the double inequality:
$$ 1\,\ll\,l\,\ll\,L \eqno (4.21)
$$
The left inequality is a macroscopic feature of the state while
the right one refers to quantum effects.
The combination of both allows, particularly, for a classical description,
at least in the standard map, of the statistical 
relaxation to the quantum steady state 
by a phenomenological diffusion equation [66 ,4]
for the Green function:
$$ \frac{\partial g(\nu,\,\sigma)}{\partial\sigma}\,=\,
   \frac{1}{4}\,
   \frac{\partial^2 g}{\partial\nu^2}\,+\,B(\nu)\,\frac{\partial g}
   {\partial\nu} \eqno (4.22)
$$
Here $g(\nu ,\,0)=|\psi (\nu ,0)|^2=\delta (\nu -\nu_0)$ and
$$ \nu\,=\,\frac{n}{2D_0}\,, \quad \sigma\,=\,\ln{(1\,+\,
   \tau_R)}\,,   \quad \tau_R\,=\,\frac{t}{2D_0} \eqno (4.23)
$$
The additional drift term in the diffusion equation with
$$ B(\nu )\,=\,{\rm sign}(\nu\,-\,\nu_0)\,=\,\pm 1 \eqno (4.24)
$$
describes the so--called quantum coherent backscattering, which is the 
dynamical mechanism of localization.

The solution of Eq.(4.22) reads [4]:
$$
  g(\nu ,\,\sigma )\,=\,\frac{1}{\sqrt{\pi\sigma}}\,\exp{\left[\,-\,
  \frac{(\delta\,+\,\sigma )^2}{\sigma}\right]}\,+\,\exp{(\,-\,4\delta )}
  \cdot{\rm erfc}\left(\frac{\delta\,-\,\sigma}{\sqrt{\sigma}}\right)
  \eqno (4.25)
$$
where $\delta =|\nu -\nu_0|$, and
$$
  {\rm erfc}(u)\,=\,\frac{2}{\sqrt{\pi}}\,\int_u^{\infty}\,{\rm e}^{-\,v^2}\,dv
$$
Asymptotically, as $\sigma\to\infty$, the Green function $g(\nu ,\,\sigma )
\to\,2\,\exp{(-4\delta )}\equiv g_s$ approaches the localized steady state
$g_s$, exponentially in $\sigma$ but only as a power--law in physical time
$\tau_R$ or $t$ ($g-g_s\sim 1/\tau_R$). This is the effect of discrete motion
spectrum. Numerical experiments confirm prediction (4.25), at least, to the
logarithmic accuracy $\sim\sigma\approx\ln{\tau_R}$ [70, 4]. 

The quantum diffusion on relaxation time scale depends, generally, on two
other conditions. The first one requires a sufficiently strong perturbation.
Otherwise, the quantum transitions between unperturbed states would be
suppressed which is called {\it perturbative localization}. 
This is a well--known
quantum effect also related to the discrete quantum spectrum. The opposite
case of strong perturbation is called {\it quasicontinuum} (referring to
the same spectrum). For the standard map this condition reads: $k\gg 1$
(see Eq.(4.13)).

The second condition is especially simple for a bounded map, e.g., $k\lsim L$
in case of standard map on a torus. This condition is required in both
quantum as well as classical systems. Otherwise, the diffusion approximation
is no longer valid, and a more complicated kinetic equation is necessary for
the description of statistical relaxation. In continuous time this condition
is formulated in terms of the dynamical time scale of the relaxation process
which the former is just one iteration of the map. The general condition
requires the dynamical change of variables to be sufficiently small.

A physical example of localization is the quantum suppression of diffusive 
photoeffect in Hydrogen atom (Lecture 3). In quantum analysis it is
convenient to change the electron energy $E$ for the number of electric field
photons: $E\to n_{\phi}=(E_0-|E|)/\Omega$ where $E_0=1/2n^2$ is the initial
energy. The quantum suppression of diffusive ionization depends on the ratio
(cf. Eq.(4.20))
$$
  \lambda_{\phi}\,=\,\frac{l_s}{n^0_{\phi}}\,\approx\,\frac{D_0}{n^0_{\phi}}\,
  \approx\,\frac{6.6\,f_n}{\Omega_n^{7/3}} \eqno (4.26)
$$
where $D_0\approx 3.3f^2/\Omega^{10/3}$ is 'classical' diffusion rate, and
$n^0_{\phi}=E_0/\Omega$ the number of absorbed photons required for 
ionization. Notice that $D_0$ does not depend on quantum number $n$, so that
the whole ionization process can be described by the local map which
considerably simplifies the theoretical analysis. 

If $\lambda_{\phi}\gsim 1$ localization does not affect the diffusion which
eventually leads to the complete ionization of the atom. For 
$\lambda_{\phi}\ll 1$ the ionization is strongly (but not completely)
suppressed due to quantum effects. Depending on parameters the suppression
may occur no matter how large is quantum number $n$. Again, this is a typical
mesoscopic phenomenon which had been predicted by the theory of quantum chaos,
and was subsequently observed in laboratory experiments (see Ref.[32]). 

The mesoscopic domain of localized quantum chaos corresponds to the interval:
$f_n^{(b)}<f_n<f_n^{(l)}$. Here $f_n^{(l)}$ is the border of localization
$\lambda_{\phi}\approx 1$ (4.26), and $f_n^{(b)}$ the chaos border (3.32). 
The size of this domain rapidly grows with $\Omega_n$:
$$
  \frac{f_n^{(l)}}{f_n^{(b)}}\,\approx\,7.6\,\Omega_n^{8/3} \eqno (4.27)
$$
The two additional conditions for quantum diffusion mentioned above lead to
the restriction: $\Omega_n\lsim n$.
  
\subsection{Examples of pseudochaos in classical mechanics}

The pseudochaos is a new generic dynamical phenomenon missed in the ergodic
theory. No doubt, the most important particular case of pseudochaos is the
quantum chaos. Nevertheless, pseudochaos occurs in classical mechanics as well.
Here are a few examples of classical pseudochaos which may help to understand
the physical nature of quantum chaos, my primary goal in these Lectures.
Besides, this unveils new features of classical dynamics as well. 

{\bf Linear waves} is the most close to quantum mechanics example of
pseudochaos (see, e.g., Ref.[71]). I remind that here only a part of quantum 
dynamics is discussed, one described, e.g., by the Schr\"odinger equation
which is a linear wave equation. For this reason the quantum chaos is called
sometime wave chaos [72]. Classical electromagnetic waves are used in
laboratory experiments as a physical model for quantum chaos [73]. The
'classical' limit corresponds here to the geometrical optics, and the
'quantum' parameter $q=L/\lambda$ is the ratio of a characteristic size $L$ 
of the system to wave length $\lambda$. As is well known in optics, no
matter how large is the ratio $\L/\lambda$ the diffraction pattern prevails at
a sufficiently far distance $R\gsim L^2/\lambda$. This is a sort of relaxation
scale: $R/\lambda\sim q^2$. 

{\bf Linear oscillator} (many--dimensional) is also a particular 
representation of waves
(without dispersion). A broad class of quantum systems can be reduced to this
model [74]. Statistical properties of linear oscillator, particularly in the
thermodynamic limit ($N\to\infty$), were studied in Ref.[75] in the frames
of TSM. On the other hand, the theory of quantum chaos suggests more rich 
behavior for a big but finite $N$, particularly, the characteristic time
scales for the harmonic oscillator motion [76], the number of freedoms $N$
playing a role of the 'quantum' parameter.

{\bf Completely integrable nonlinear systems} also reveal pseudochaotic 
behavior. An example of statistical relaxation in the Toda lattice had been
presented in Ref.[77] much before the problem of quantum chaos arose. Moreover,
the strongest statistical properties in the limit $N\to\infty$, including one
equivalent to the exponential instability (the so--called $K$--property) were
rigorously proved just for the (infinite) completely integrable systems (see
Ref.[10]). 

{\bf Digital computer} is a very specific classical dynamical system whose
dynamics is extremely important in view of the ever growing interest to
numerical experiments covering now all branches of science and beyond.
The computer is the 'overquantized' system in that {\it any} quantity here is
{\it discrete} while in quantum mechanics only the product of two conjugated
variables does so. 'Quantum' parameter here $q=M$ which is the largest computer
integer, and the short time scale (4.5) $t_r\sim\ln{M}$ which is the number of
digits in the computer word[51]. Owing to the discreteness, any dynamical
trajectory in computer becomes eventually periodic, the effect well known 
in the theory and practice of the so--called pseudorandom number generators.
The term 'pseudochaos' itself was borrowed from just this particular
example [68, 4].
One should take all necessary precautions to exclude this computer artifact
in numerical experiments (see, e.g., [78] and references therein). 
On the mathematical part, the periodic approximations
in dynamical systems are also studied in the ergodic theory, apparently
without any relation to pseudochaos in quantum mechanics or computer [10].

The computer pseudochaos is the best answer to ones who
refuse accept the quantum chaos as, at least, a kind of chaos,
and who still insist that only the classical--like (asymptotic) chaos deserves
this name, the same chaos which was (and is) studied to a large extent just
on computer that is the chaos inferred from a pseudochaos!

\section{Conclusion: old challenges and new hopes}

The discovery and understanding of the new surprising phenomenon -- dynamical
chaos -- opened up new horizons in solving many other problems including some
long--standing ones. Here I can give only a 
preliminary consideration of possible new approaches to such problems together
with some plausible conjectures (see also Ref.[4]).

Let us begin with the problem directly related to quantum dynamics, namely,
the quantum measurement or, to be more correct, the specific stage of the
latter, the {\it $\psi$--collapse}. It is just the part of quantum dynamics
I bypassed above.
This part still remains very vague to the extent that there is no 
common agreement even 
on the question whether it is a real physical problem or an 
ill--posed one so that the Copenhagen interpretation of 
(or convention in) quantum mechanics gives 
satisfactory answers to all the {\it admissible} questions.
In any event, there exists as yet no dynamical description of the 
quantum measurement including $\psi$--collapse.
The quantum measurement, as far as the result is concerned, is 
commonly understood as a
fundamentally random process. 
However, there are good reasons to hope that this randomness can be 
interpreted as a particular manifestation of dynamical chaos [79].

The Copenhagen convention was (and still remains) very important as a
phenomenological link between a very specific quantum theory and the laboratory
experiments. Without this link the studies of microworld would be simply
impossible. The Copenhagen philosophy perfectly matches the standard 
experimental setup of two measurements: the first one fixes the initial
quantum state, and the second records the changes in the system. However,
it is less clear how to deal with {\it natural processes} without any
man--made measurements that is without notorious {\it observer}. Since the
beginning of quantum mechanics such a question has been considered ill--posed
(meaning nasty). However, now there is a revival of interest to a deeper 
insight into this problem (see, e.g., Ref.[79]). Particularly, Gell-Mann and
Hartle put a similar question, true, in the context of a very specific and
global problem -- the quantum birth of the Universe [80]. In my understanding,
such a question arises as well in much simpler problems concerning any
natural quantum processes. What is more important, the answer [80] does not
seem to be satisfactory. Essentially, it is the substitution of the automaton
(Information gathering and utilizing system) for the standard human observer.
Neither seems to be a generic construction in the microworld. 

The theory of quantum chaos allows us to solve, at least, the (simpler) half
of the $\psi$--collapse problem. 
Indeed, the measurement device is by purpose a macroscopic 
system for which the 
classical description is a very good approximation. In such a 
system the strong chaos with 
exponential instability is quite possible. The chaos in the 
measurement classical device is not 
only possible but unavoidable since the measurement system has to be, 
by purpose again, a highly 
unstable system where a microscopic intervention produces the 
macroscopic effect. 
The importance of chaos for the quantum measurement is in that it 
destroys the coherence of the 
initial pure quantum state to be measured converting it into 
the incoherent mixture. In the 
present theories of quantum measurement this is described as 
the effect of the external noise (see, e.g., Ref.[81]).
True, the noise is sufficient to destroy the quantum coherence, yet
it is not necessary at all [82]. 
The chaos theory allows to get rid of the unsatisfactory effect 
of the external noise and to 
develop a purely dynamical theory for the loss of quantum coherence.
Yet, this is not the whole story. If we are satisfied with
the {\it statistical} description of quantum dynamics (measurement including)
then the decoherence is all we need. However, the {\it individual} behavior
includes the second (main) part of $\psi$--collapse, namely, the 
{\it concentration} of $\psi$ in a single state of the original superposition
$$ \psi\,=\,\sum_n\,c_n\,\psi_n\,\to\,\psi_k, \qquad \sum_n\,|c_n|^2\,=\,1
$$
This is the proper $\psi$--collapse to be understood.

Also, it is another challenge to the correspondence principle. For the
quantum mechanics to be universal it must explain as well the very specific
classical phenomenon of the {\it event} which does happen and remains for ever
in the classical records, and which is completely foreign to the proper 
quantum mechanics. It is just the effect of $\psi$--collapse.

All these problems could be resolved by a hypothetical phenomenon of
{\it selfcollapse} that is the collapse without any 'observer', human or 
automatic. 
Recently, some attempts to resolve this latter problem were made[83] which are
still to be understood and evaluated. So far I would like simply to mention
that these attempts are trying to make use of the nonlinear ``semiquantum''
equations like the well studied nonlinear Schr\"odinger equation (for
discussion see Refs.[4, 84]). 

Now we come to even more difficult problem of the {\it causality principle}
that is the universal time ordering of the events. This principle has been
well confirmed by numerous experiments in all branches of physics. It is
frequently used in the construction of various theories but, to my knowledge,
any general relation of causality to the rest of physics was never studied.

This principle looks as a statistical law (another time arrow), hence a new
hope to understand the mechanism of causality via dynamical chaos. Yet, it
directly enters the dynamics as the additional constraint on the interaction
and/or the solutions of dynamical equations. A well known and quite general
example is in keeping the retarded solutions of a wave equation only
discarding advanced ones as 'nonphysical'. However, this is generally 
impossible because of the boundary conditions. Still,
the causality holds true as well. 

In some simple classical {\it dissipative} models like a driven damping
oscillator the dissipation was shown to imply causality [85, 86]. However,
such results were formulated as the restriction on a class of systems showing
causality rather than the foundations of the causality principle. Nevertheless,
it was already some indication on a possible physical connection between
dynamical causality and statistical behavior. To my knowledge, this connection
was never studied farther. To the contrary, the development of the theory
went the opposite way: taking for granted the causality  to deduce all possible
consequences, particularly, various dispersion relations [86]. 
In some physical [87] (not mathematical [10]!) theories in TSM the causality
principle, modestly termed sometimes as 'causality condition', 
is used to 'derive'
statistical irreversibility from the time--reversible dynamics. As was
discussed in Lecture 2 the physical chaos theory (and, implicitly, the
mathematical ergodic theory as well) predicts, instead, the nonrecurrent
relaxation without any additional 'conditions', causality including.
Then, the above--mentioned arguments (e.g., in Ref.[87]) could be reversed
in such a way to derive the causality from the dynamical chaos, similar to
Refs.[85, 86] but for a much more general class of dynamical systems.

The causality relates two qualitatively different kinds of events: {\it causes}
and {\it effects}. The former may be simply the initial conditions of motion,
the point missed in the above--mentioned examples of causality--dissipation
relation. The initial conditions not only formally fix a particular trajectory
but they are also {\it arbitrary} which is, perhaps, the key point in the 
causality problem. Also, this may shed some light on another puzzling 
peculiarity of {\it all} known dynamical laws: they describe the motion up to
arbitrary initial conditions only. It looks like the 
dynamical laws include
already the causality implicitly even though they do not this explicitly. 
In any event, something arbitrary suggests a chaos around.

Again, we arrive at a tangle of interrelated problems. A plausible conjecture
how to resolve them might be as follows. Arbitrary cause indicates some
statistical behavior while the cause--effect relation points out a dynamical
law. Then, we may conjecture that when the cause acts the transition from
statistical to dynamical behavior occurs which separates statistically the
cause from the 'past' and fixes dynamically the effect in the 'future'.
In this imaginary picture the 'past' and 'future' are related not to the time
but rather to cause and effect, respectively. Thus, the causality might be 
not the time ordering (time arrow) but {\it cause--effect ordering}, 
or {\it causality
arrow}. The latter is very similar to the process arrow, discussed in Leture 2,
both always pointing in the same direction.
Now, the central point is in that the cause is arbitrary while the effect is
not whatever the time ordering.

This is, of course, but a raw guess to be developed, carefully analysed, and
eventually confirmed or disproved experimentally. 

Also, this picture seems to be closer to the statistical (secondary) dynamics
(synergetics, or $S\supset D$ inclusion in (1.1)) rather than to dynamical
chaos. Does it mean that the primary physical laws are statistical or, instead,
that the chain of inclusions (1.1) is actually a closed ring with a 'feedback'
coupling the secondary statistics to the primary dynamics?

We don't know.

In these Lectures I has never given the definition of dynamical chaos,
either classical or quantal, restricting myself to informal explanations (see
Ref.[4] for some current definitions of chaos).
In a mathematical theory the definition of the main object of the 
theory precedes the results; in 
physics, especially in new fields, it is quite often vice versa.
First, one studies a new phenomenon like dynamical chaos and only at 
a later  stage, after 
understanding it sufficiently, we try to classify it, to find its 
proper place in the existing theories 
and eventually to choose the most reasonable definition.

This time has not yet come.

\end{document}